\documentclass[12pt,preprint]{aastex}
\usepackage{emulateapj5}

\def\ast{\mathchar"2203} \mathcode`*="002A 
  
\def\=={{\equiv}}

\def\mdot{\dot M}

\def\Lbol{{L_{bol}}} 
 
\def\Rstar{{\relax \ifmmode {R_\ast} \else $R_\ast$\fi}} 
\def\rstar{{\relax \ifmmode {R_\ast} \else $R_\ast$\fi}} 
\def\rosat{{\it ROSAT\,}}

\def \asca{{\it ASCA}}

\def \zpup{\hbox{$\zeta$ Pup}}
\def \zori{\hbox{$\zeta$ Ori}}
\def \fx{\relax \ifmmode {f_{\rm X}} \else $f_{\rm X}$\fi}
\def \Lx{\relax \ifmmode {L_{\rm X}} \else $L_{\rm X}$\fi}
\def \lx{\relax \ifmmode {L_{\rm X}} \else $L_{\rm X}$\fi}
\def \lxlb{\relax \ifmmode {L_{\rm X}/L_{\rm Bol}} \else $L_{\rm
    X}/L_{\rm Bol}$\fi} 
\def \Lbol{\relax\ifmmode{L_{\rm Bol}}\else $L_{\rm Bol}$\fi}
\def \lb{\relax\ifmmode{L_{\rm Bol}}\else $L_{\rm Bol}$\fi}

\def\msunyr{\hbox{${\rm M}_\odot\,$  yr $^{-1}$}}
\def \taunu{\relax \ifmmode {\tau_{\nu}} \else $\tau_{\nu}$ \fi}
\def \sigmanu{\relax \ifmmode {\sigma_{\nu}} \else $\sigma_{\nu}$ \fi}

\def \einstein{{\it Einstein}}

\def \chandra{{\it Chandra}}
\def \xmm{{\it XMM}}

\def \etal{et~al.}

\def \mdovervinf{\relax \ifmmode {\dot M}/v_{\infty} \else ${\dot M}/v_{\infty}$\fi}
 
\def\ltwig{\mathrel{\spose{\lower 3pt\hbox{$\mathchar"218$}} 
     \raise 2.0pt\hbox{$\mathchar"13C$}}} 
\def\gtwig{\mathrel{\spose{\lower 3pt\hbox{$\mathchar"218$}}  
     \raise 2.0pt\hbox{$\mathchar"13E$}}} 
\def\spose#1{\hbox to 0pt{#1\hss}} 
 
\def \tsco{\hbox{$\tau$ Sco}}
\def \t1oric{\hbox{${\theta}^1$ Ori C}}
\def \tori{\hbox{${\theta}^1$ Ori C}}
\def\kms{\hbox{km s$^{-1}$}}
\def \hetgs{HETGS}

\def \f2i{{\it f/i}}
\def \ftoi{{\it f/i}}

\shorttitle{Chandra Spectroscopy of tau Sco}
\shortauthors{Cohen et al.}

\begin{document}

\title{High-Resolution {\it CHANDRA} Spectroscopy of tau Scorpii: A
  Narrow-Line X-ray Spectrum From a Hot Star}

\author{David H. Cohen and Genevi\`{e}ve E. de Messi\`{e}res}
\affil{Department of Physics and Astronomy, Swarthmore College,
  Swarthmore, PA 19081} \email{cohen@astro.swarthmore.edu,
  gdemess1@swarthmore.edu}

\author{Joseph J. MacFarlane} \affil{Prism Computational Sciences, 455
  Science Dr., Madison, WI 53711} \email{jjm@prism-cs.com}

\author{Nathan A. Miller} \affil{Department of Physics and Astronomy,
  105 Garfield Ave., University of Wisconsin, Eau Claire, Eau Claire,
  WI 54702} \email{millerna@uwec.edu}

\author{Joseph P. Cassinelli} \affil{Department of Astronomy, 475 N.
  Charter St., University of Wisconsin, Madison, Madison, WI 53706}
\email{cassinelli@astro.wisc.edu}

\author{Stanley P. Owocki} \affil{Bartol Research Institute,
  University of Delaware, Newark, DE 19716}
\email{owocki@bartol.udel.edu}

\author{and \\ Duane A. Liedahl}
\affil{Lawrence Livermore National Laboratory, Livermore, CA 94550}
\email{liedahl1@llnl.gov}

\begin{abstract}
  
  Long known to be an unusual early-type star by virtue of its hard
  and strong X-ray emission, $\tau$ Scorpii poses a severe challenge
  to the standard picture of O-star wind-shock X-ray emission. The
  \chandra\/ \hetgs\/ spectrum now provides significant direct
  evidence that this B0.2 star does not fit this standard wind-shock
  framework. The many emission lines detected with the \chandra\/
  gratings are significantly narrower than what would be expected from
  a star with the known wind properties of \tsco, although they are
  broader than the corresponding lines seen in late-type coronal
  sources.  While line ratios are consistent with the hot plasma on
  this star being within a few stellar radii of the photosphere, from
  at least one He-like complex there is evidence that the X-ray
  emitting plasma is located more than a stellar radius above the
  photosphere. The \chandra\/ spectrum of \tsco\/ is harder and more
  variable than those of other hot stars, with the exception of the
  young magnetized O star \tori.  We discuss these new results in the
  context of wind, coronal, and hybrid wind-magnetic models of
  hot-star X-ray emission.
  
\end{abstract}

\keywords{circumstellar matter -- stars: early-type -- stars: mass
  loss -- X-rays: stars}

\section{Introduction}

The surprising discovery of X-ray emission from OB stars by the
\einstein\/ satellite in the late 1970s (Harnden \etal\ 1979; and
anticipated by UV superionization observations [Cassinelli \& Olson
1979]) is now a 25-year-old mystery.  Because magnetic dynamos and the
associated coronal X-rays were not believed to exist in hot stars,
alternative models have been sought since the discovery of X-rays from
hot stars. Most of these center on some type of wind-shock mechanism
that taps the copious energy in the massive radiation-driven winds of
hot stars (Lucy \& Solomon 1970; Castor, Abbott, \& Klein 1975,
hereafter CAK).  Wind-shock models, especially the line-force
instability shock model (Owocki, Castor, \& Rybicki 1988; Feldmeier,
Puls, \& Pauldrach 1997; Feldmeier \etal\ 1997), have become the
favored explanations for the observed OB star X-ray properties.

However, various pieces of evidence have recently emerged indicating
that magnetic fields exist on some hot stars (Henrichs \etal\ 2000;
Donati \etal\ 2001, 2002).  There is also mounting evidence that the
standard wind-shock scenario is inadequate for explaining the observed
levels of X-ray production in many B stars (Bergh\"{o}fer \& Schmitt
1994; Cohen, Cassinelli, \& MacFarlane 1997a; Cohen, Cassinelli, \&
Waldron 1997b), and perhaps in some O stars as well (Schulz \etal\/
2001; Waldron \& Cassinelli 2001; Miller \etal\/ 2002), leading to a
revival in the idea that some type of dynamo mechanism might be
operating on hot stars.  Theoretical investigations of magnetized OB
stars have followed (Charbonneau \& MacGregor 2001; MacGregor \&
Cassinelli 2002), including studies of hybrid wind-magnetic models
(Babel \& Montmerle 1997; ud-Doula \& Owocki 2002; Cassinelli \etal\/
2002).

Discriminating between the two broad pictures of X-ray emission in hot
stars (magnetic coronal vs. wind shock) has been very difficult during
the era of low-resolution X-ray spectroscopy. But with the launch of
\chandra\/ in 1999, a new, high-resolution window has opened, and we
can now apply quantitative diagnostics in our attempts to understand
the high-energy mechanisms operating in the most massive and luminous
stars in our galaxy.

Coronal heating mechanisms are complex and, even in our own Sun, not
fully understood.  Their possible application to hot stars is
certainly not clear, even in those hot stars for which magnetic fields
have been detected.  However, we can assume that any type of coronal
X-ray-production mechanism involves the confinement of relatively
stationary hot plasma near the photosphere.  This is in contrast to
wind shocks, which, whatever their nature (and many different specific
models have been proposed), involve low-density hot plasma farther
from the photosphere and moving at an appreciable velocity.  This
broad dichotomy naturally leads to several possible observational
discriminants that might be implemented using high-resolution X-ray
spectroscopy.  These include the analysis of line widths (or even
profiles), which could reflect Doppler broadening from the bulk
outflow of winds (which in normal OB stars have terminal velocities of
up to 3000 km s$^{-1}$) and emission-line ratios that are sensitive to
density, the local radiation field, or (unfortunately) in some cases,
both.  There are numerous other spectral diagnostics available,
including temperature-sensitive line ratios, spectral signatures of
wind absorption, time variability of individual lines, and
differential emission measure (DEM) fitting (multiline plus continuum
global spectral modeling of X-ray emission from plasma with a
continuous distribution of temperatures).  All of these can be useful
for determining the physical properties of the hot plasma on OB stars,
and, ultimately for constraining models, but none can provide the same
level of direct discrimination between the wind-shock and coronal
scenarios as line profiles and ratios.

In this paper we report on the very high signal-to-noise,
high-resolution \chandra\/ \hetgs\/ spectrum of the B0.2 V star \tsco.
We chose to observe this particular hot star because it is one of the
most unusual single hot stars, with indications already from various
low-resolution X-ray observations that its X-ray properties are
extreme and present a severe test to the standard wind-shock scenario.
As such, it makes an interesting contrast to the five single O stars
that have been observed at high resolution with \chandra\/ and \xmm\/
up to this point.  In fact, some of the unusual properties of \tsco\/
(discussed below in \S3) have already inspired the development of
alternate models of X-ray production, notably including the ``clump
infall'' model of Howk \etal\/ (2000).  In this scenario overdense
regions condense out of the wind, and the reduced line force due to
the increased optical depth in the clumps results in these clumps
falling back toward the star, generating strong shocks via
interactions with the outflowing mean wind.

\section{Observations}

The data presented in this paper were taken between 2000 September 17
and 18, over two different \chandra\/ orbits, with an effective total
exposure time of 72 ks.  We used the ACIS-S/HETGS configuration and
report on the medium-energy grating (MEG) and high-energy grating
(HEG) spectra, although we concentrate primarily on the MEG 1st order
spectrum, which contains a total of 19348 counts (while the HEG 1st
order spectrum has 5230 total counts).  We extracted and analyzed the
spectra with CIAO tools (Version 2.1.2), including the SHERPA
model-fitting engine. We used the CALDB libraries, Version 2.6, and
the OSIP files from 2000 August to build auxiliary response files and
grating response matrices.  We fit line-profile models using SHERPA,
but astrophysical modeling of the spectral data took place outside of
the CIAO environment with a custom non-LTE spectral synthesis code,
described in section 5.2.

\section{The Star, \tsco}

A member of the upper Sco-Cen association and a Morgan and Keenan
system standard, the B0.2 V star \tsco\/ has been the subject of much
study over the past several decades.  At a distance of $132 \pm 14$ pc
(Perryman \etal\/ 1997), this $m_v=2.8$ star is closer to the Earth
than any hotter star.  It has an effective temperature of $31,400 \pm
300$ K and ${\rm{log}}~ g=4.24 \pm .03$ (Kilian 1992).  UV
observations show evidence for a wind with a terminal velocity of
roughly 2000 \kms\/ (Lamers \& Rogerson 1978).  The mass-loss rate is
difficult to determine because of the uncertain ionization correction,
but it seems consistent with the theoretically expected value of
$\mdot \approx 10^{-8}$ \msunyr\/ (calculations based on Abbott 1982).
Published values based on UV absorption lines (e.g., Lamers \&
Rogerson 1978) and a recently determined upper limit based on IR
emission lines (Zaal \etal\/ 1999) are roughly half this value.  These
wind properties contrast with those of the prototypical early O
supergiant \zpup, which has a mass-loss rate at least 100 times larger
(because of its significantly larger luminosity).  \chandra\/ \hetgs\/
observations of \zpup\/ were recently reported by Cassinelli \etal\/
(2001), and clear signatures of the stellar wind (broadening and
absorption effects) are seen in those data.

There are several unusual properties of \tsco\/ that might be relevant
for our attempts to understand its X-ray properties: (1) It has a
quite highly ionized wind, since the star is the coolest to show
\ion{O}{6} in its {\it COPERNICUS} spectrum (Lamers \& Rogerson 1978).
(2) The wind is also stronger than normal for the star's spectral type
(Walborn, Parker, \& Nichols 1995). (3) It has a very low $v_{rot}{\rm
  sin}i$ of $10 \pm 5$ \kms\/ (Wolff, Edwards, \& Preston 1982). (4)
It shows unusual redshifted absorption of about 200 \kms\/ in some of
the higher ionization UV lines (which also show much more extended
blue absorption, as one would expect; Lamers \& Rogerson 1978). (5) It
is extremely young; perhaps being only 1 million years on the main
sequence (Kilian 1994).  (6) It shows unusually strong photospheric
turbulence, as seen in optical and UV absorption lines (Smith \& Karp
1979) and in the IR (Zaal \etal\ 1999).

In addition to these properties, it is already known from \einstein,
\rosat, and \asca\/ measurements that \tsco\/ has an unusually hard
X-ray spectrum and a high X-ray luminosity (Cohen \etal\ 1997a,
1997b).  It is this fact, along with the redshifted UV absorption
(item 4 above), that prompted Howk \etal\/ (2000) to develop their
model of infalling clumps for \tsco.

Many of these properties have been cited as evidence that \tsco\/ may
not fit into the usual wind-shock picture of hot star X-ray
production, or at least that its wind emission mechanism may be
modified by the presence of a magnetic field or perhaps involve some
other, less standard, wind-shock model.  We address these issues
in detail in the next sections.

\section{The Data}

\begin{figure}
\includegraphics[angle=90,scale=0.28]{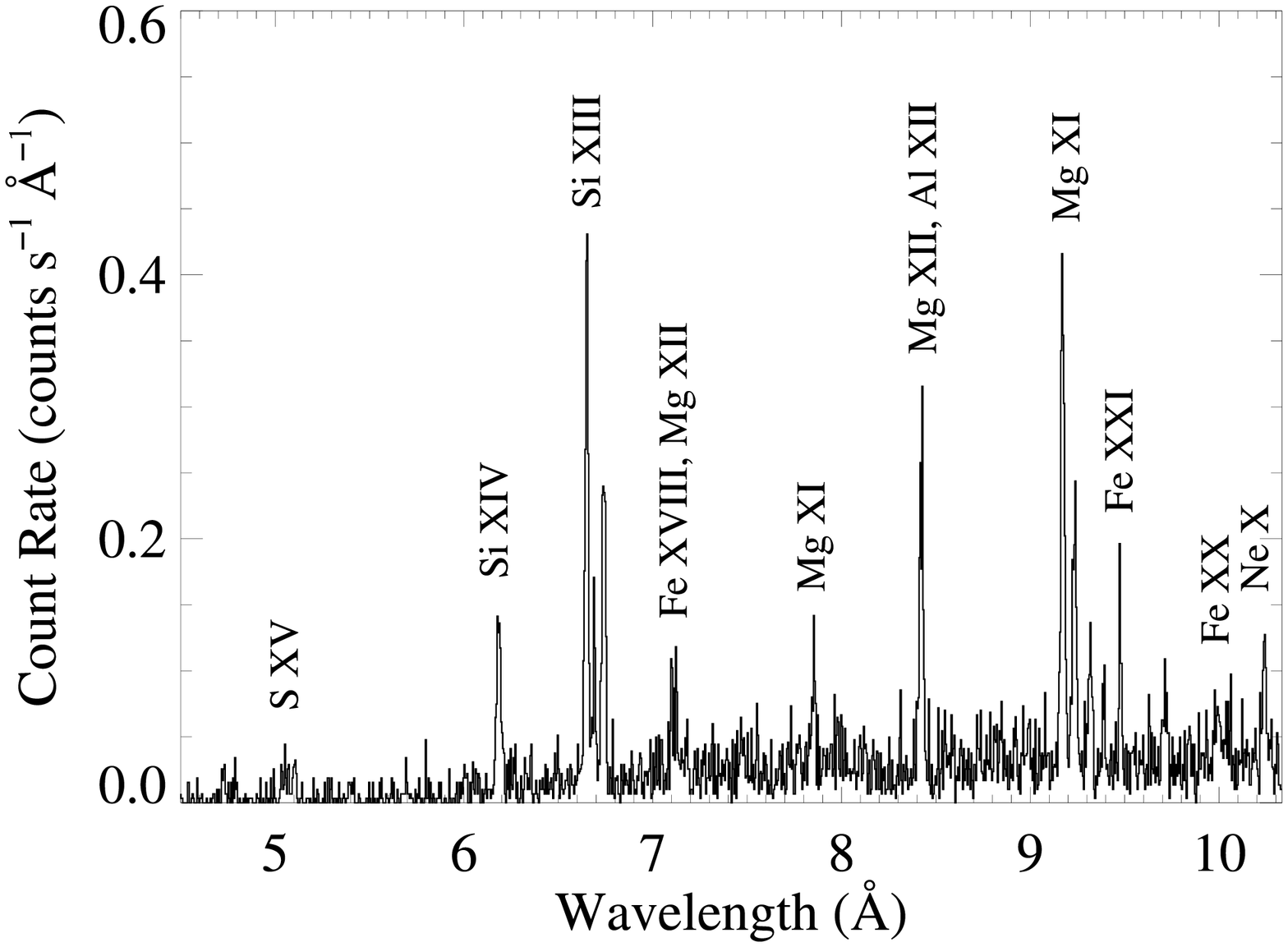}
\includegraphics[angle=90,scale=0.28]{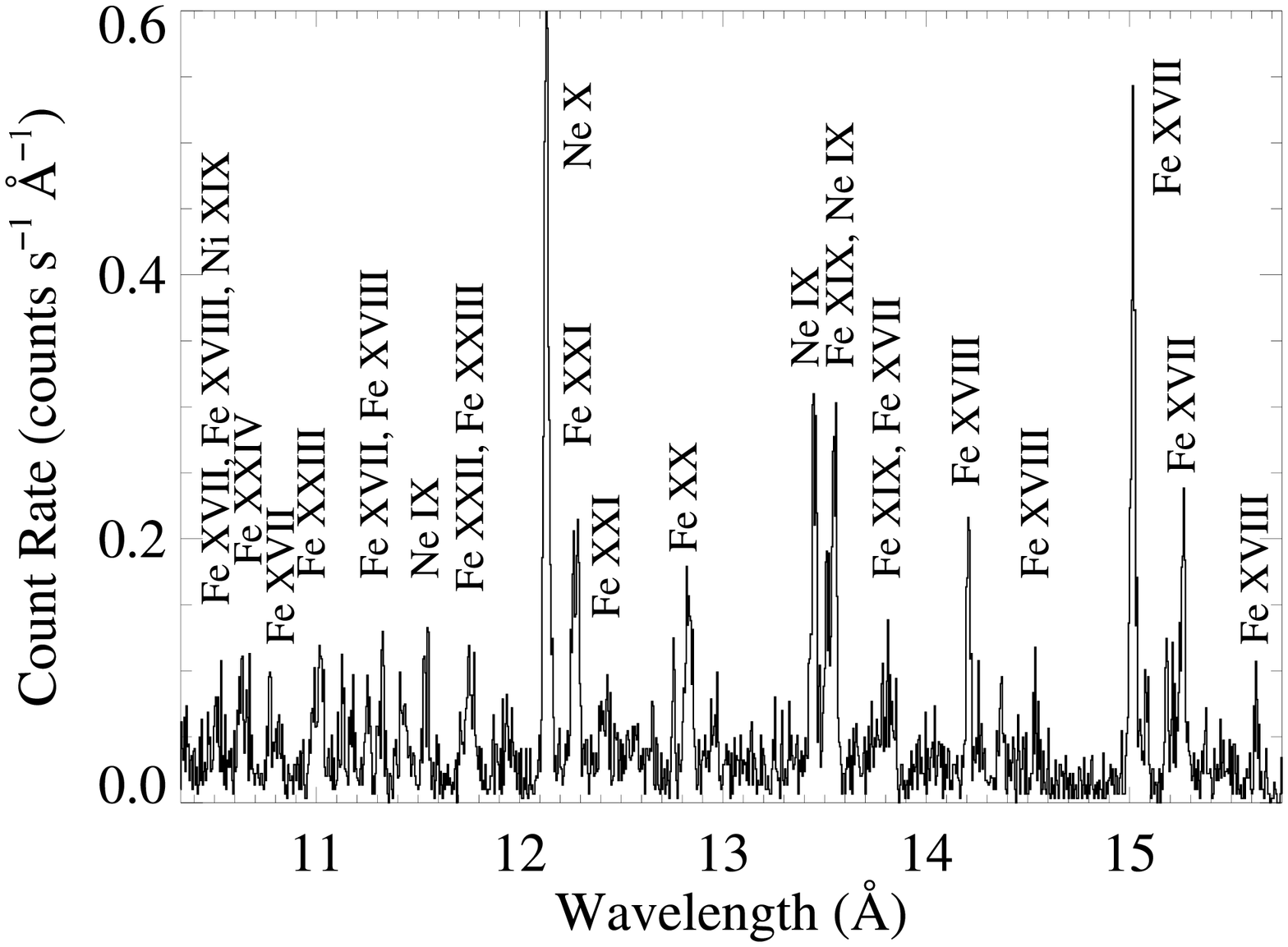}
\includegraphics[angle=90,scale=0.28]{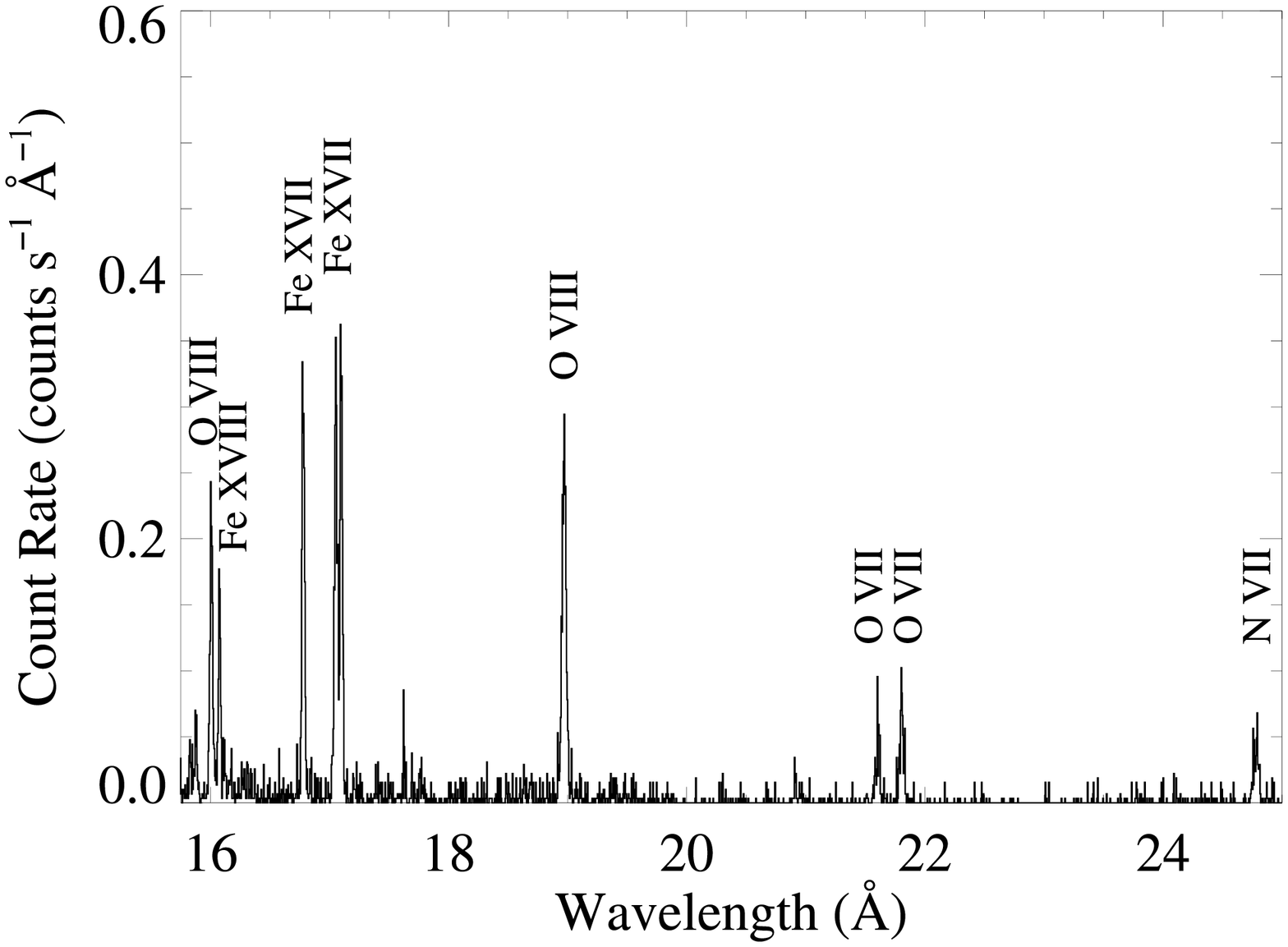}
\caption{Combined MEG +1 and -1 order spectrum of \tsco\/ with
  lines labeled}
\label{fig:spec_main}
\end{figure}

\begin{figure}
\begin{center}
\includegraphics[angle=90,scale=0.7]{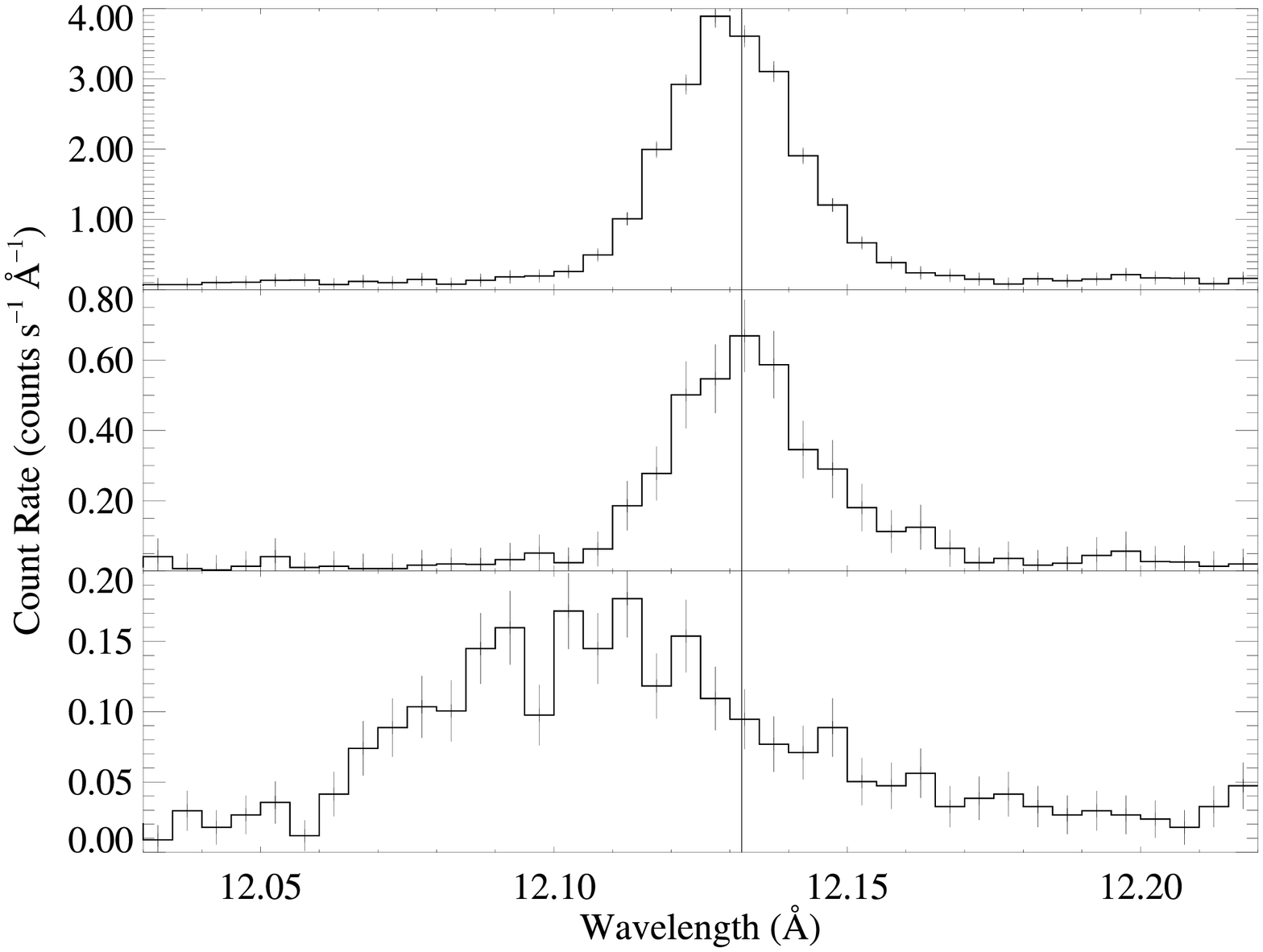}
\end{center}
\caption{Combined MEG +1 and -1 order spectra of Capella, \tsco,
  and \zpup\/ ({\it top to bottom}), centered on the region around the
  \ion{Ne}{10} line at 12.13 \AA.}
\label{fig:spec_comp}
\end{figure}

The overall quality and richness of the \chandra\ \hetgs\/ data can be
seen in the MEG first-order spectrum shown in Figure
\ref{fig:spec_main}.  There are numerous lines present in this high
signal-to-noise spectrum, and they are much better separated than in
any of the other hot stars thus far observed (because of the small wind
broadening seen in \tsco\/ compared to these stars).  In Figure
\ref{fig:spec_comp} we show a portion of the MEG spectrum, centered on
the neon Ly$\alpha$ line, compared with data from \zpup\/ and Capella,
a typical O supergiant wind source and a late-type coronal source,
respectively.  It is immediately obvious that \tsco\/ is superficially
more similar to Capella than to \zpup.



To quantify the properties of the spectrum, we fitted every strong
line with a simple model (Gaussian plus a polynomial for the nearby
continuum), providing values of the line flux, characteristic width,
and central wavelength.  We employed the appropriate spectral response
matrices in these fits, so the derived line widths are intrinsic and
do {\it not} include instrumental broadening. When there was
sufficient signal, we fit the MEG and HEG data simultaneously.  These
results are summarized in Table \ref{tab:lines}, with the several
strongly blended spectral features listed in Table \ref{tab:blends}.
The astrophysical analysis of these data is presented in \S5.

\begin{figure}
\begin{center}
\includegraphics[angle=90,scale=0.7]{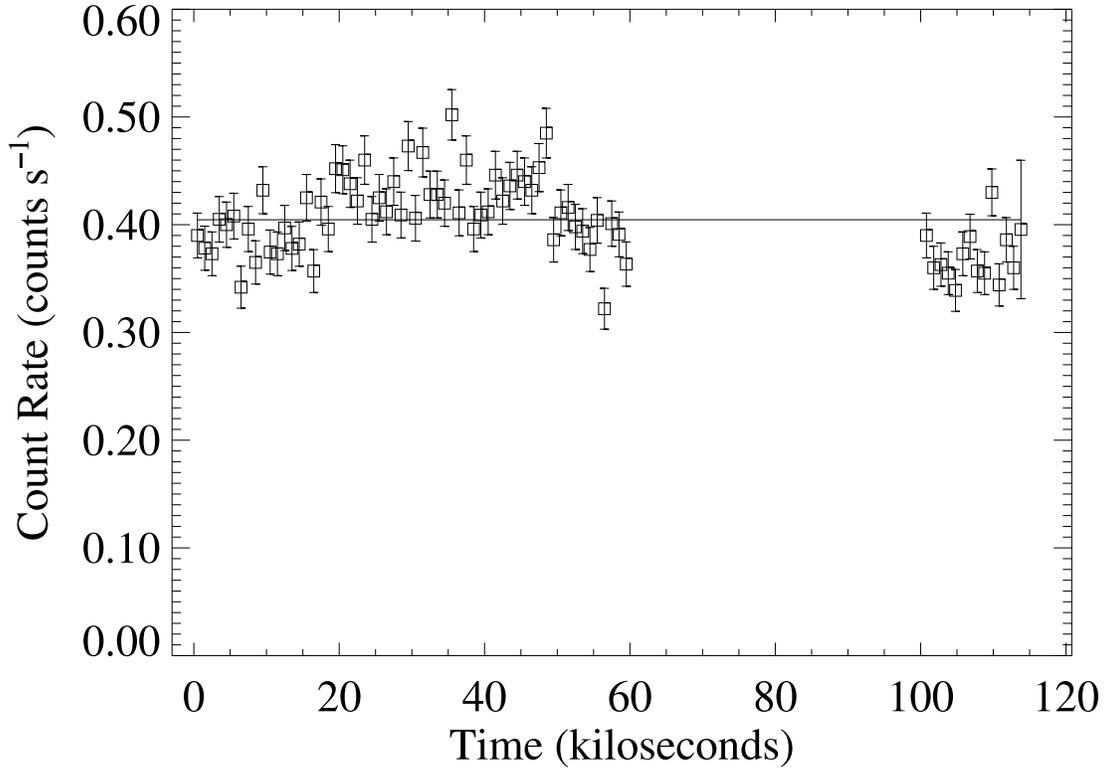}
\end{center}
\caption{X-ray light curve formed from the combined MEG +1 and -1 order
  counts, with 1000 second bins. The mean count rate is indicated by
  the dashed line.  The hypothesis of a constant source
  can be rejected at a more than 99.99\% confidence level.} 
\label{fig:lightcurve_main}
\end{figure}

\begin{figure}
\begin{center}
\includegraphics[angle=90,scale=0.7]{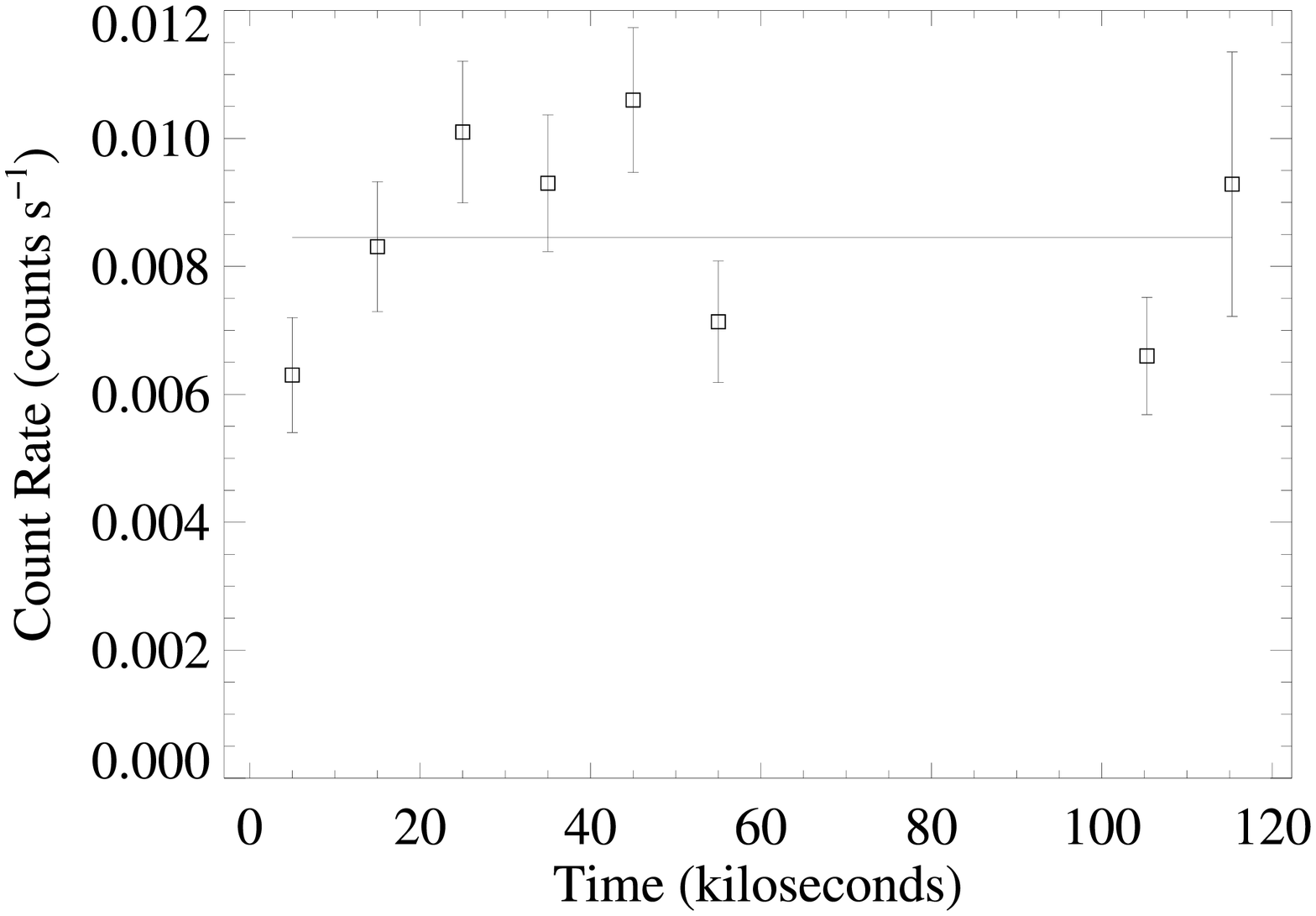}
\end{center}
\caption{X-ray light curve formed from the combined MEG +1 and -1 order
  counts in the \ion{O}{8} line at 18.97 \AA, with 10,000 s bins.  The
  mean count rate is indicated by the dashed line.  The hypothesis of
  a constant source can be rejected at a more than 99\% confidence
  level.}
\label{fig:lightcurve_oxygen}
\end{figure}

Time variability properties of \tsco\/ are also of interest in
discriminating between the two primary X-ray-production mechanisms.
In Figure \ref{fig:lightcurve_main} we show a lightcurve formed of all
the photons in the -1,-2,-3,+3,+2,+1 orders of both the MEG and the HEG.
The bins are 1000 seconds in length.  The data are not consistent with
a constant source (at $P>99.99$ \%), but no large flare-type events,
with a sudden increase in flux followed by a gradual decline back to
the basal level, are seen. The standard deviation of the data is
slightly less than 10 \% of the mean.  Only one individual line
feature, the oxygen Ly$\alpha$ line near 19 \AA, shows evidence of
significant variability. Its lightcurve is shown in Figure
\ref{fig:lightcurve_oxygen}. We tested other strong lines, and only
the \ion{Fe}{17} line at 15.01 \AA\/ shows even marginal variability.



\section{The X-ray Diagnostics and Analysis}

\subsection{Line Widths}

The emission lines of \tsco\/ are obviously narrow compared to the
same lines observed in hot-star wind X-ray sources like \zpup, but how
intrinsically narrow are they?  Might they be no broader than thermal
broadening would dictate?  To answer these questions, we fitted
Gaussian line models, convolved with the instrumental response
function, so the width of the model Gaussian reflects the actual width
of the emitted line.  We do not know what actual line profile shape is
warranted, but because the Gaussian model is simple and relatively
general and the observed line broadening is not dramatic, we did not
attempt to fit other profile shapes.

\begin{figure}
\begin{center}
\includegraphics[angle=90,scale=0.7]{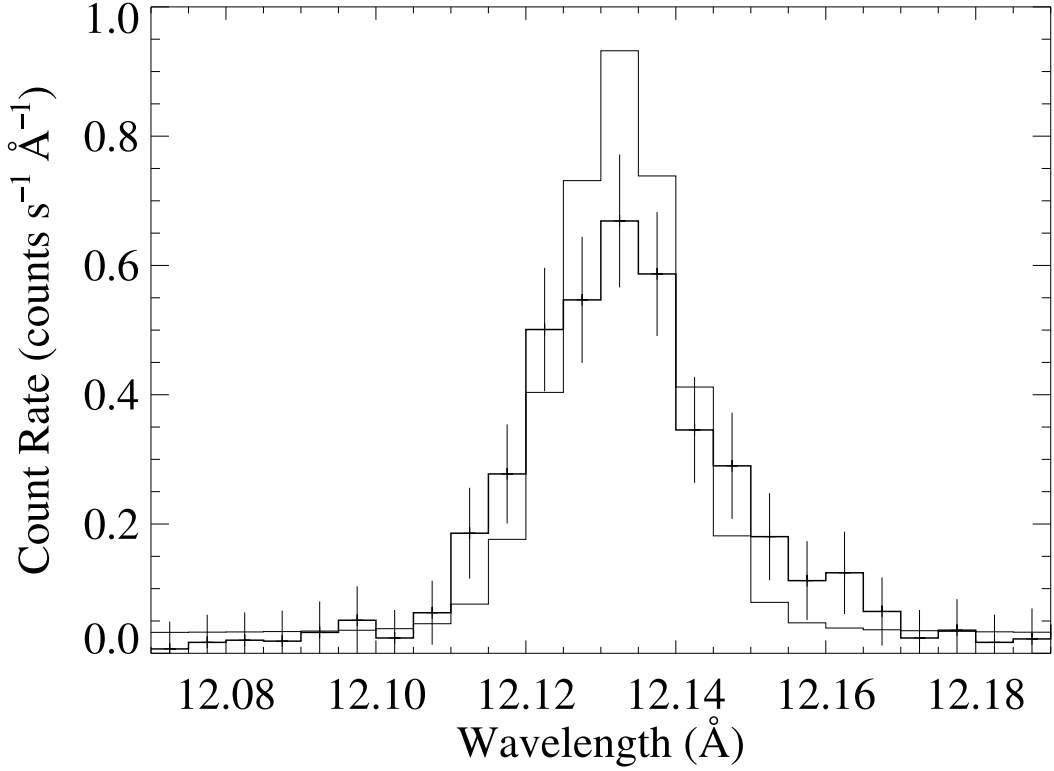}
\end{center}
\caption{MEG +1 and -1 order observation of the neon Ly$\alpha$
  line ({\it histogram}) with an intrinsically narrow model (convolved with
  the instrument response).  The fit to the data show a statistically
  significant line width.}
\label{fig:NeX_width}
\end{figure}

The line widths--half-width at half maximum (HWHM)--listed in Table
\ref{tab:lines} generally exceed by a factor of about 3 those expected
from thermal broadening (assuming a temperature equal to that at which
the emissivity for a given line peaks).  In addition, fits we
performed to Capella and AB Dor (both late-type coronal sources) using
the same procedures found narrower lines (consistent with thermal
broadening) than we found for \tsco.  In Figure \ref{fig:NeX_width} we
show the \ion{Ne}{10} line of \tsco\/ in the MEG, along with an
intrinsically narrow line (Gaussian model with $\sigma = 0$; i.e. a
delta function) convolved with the instrumental response.  The data
clearly are broader than the model.  We also note that some of the
features in the spectrum are doublets.  We generally fitted these with
single-Gaussian models.  In each case in which we fitted a doublet with a
single-line model, the separation of the two components of the doublet
was far less than the typical (and derived) line width.  Thus the
derived line widths using this procedure are only marginally elevated
because of the blended doublet.



\begin{figure}
\begin{center}
\includegraphics[angle=270,scale=0.6]{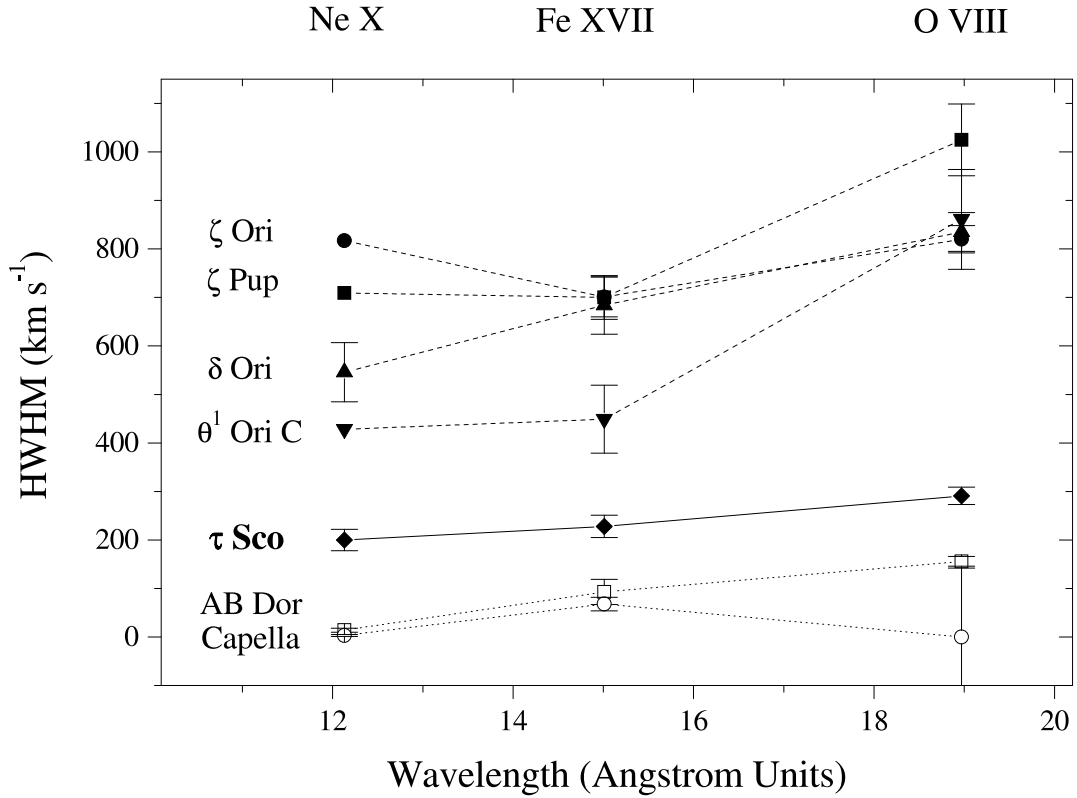}
\end{center}
\caption{Derived line widths (HWHM) for three strong lines in
  seven stars: two stars representative of coronal sources (Capella
  and AB Dor: {\it open symbols connected by dotted lines}), \tsco\/
  ({\it filled diamonds and solid line}), and four O stars ({\it
    filled symbols and dashed lines}), which are presumably wind X-ray
  sources.}
\label{fig:line_width_comparison}
\end{figure}

The velocities of most of the lines in the \hetgs\/ spectrum of
\tsco\/ are a few hundred \kms\/ (HWHM), which, while greater than the
thermal velocity, is significantly less than the wind terminal
velocity.  The X-ray line velocities seen in \tsco\/ are thus larger
than those in coronal sources and smaller than those in wind sources,
a trend that is shown in Figure \ref{fig:line_width_comparison}, and
importantly, the velocities derived for the wind of \tsco\/ are
significantly less than what would be seen if the X-ray-emitting gas
had the same velocity distribution as the UV-absorbing gas.

\begin{figure}
\begin{center}
\includegraphics[angle=270,scale=0.7]{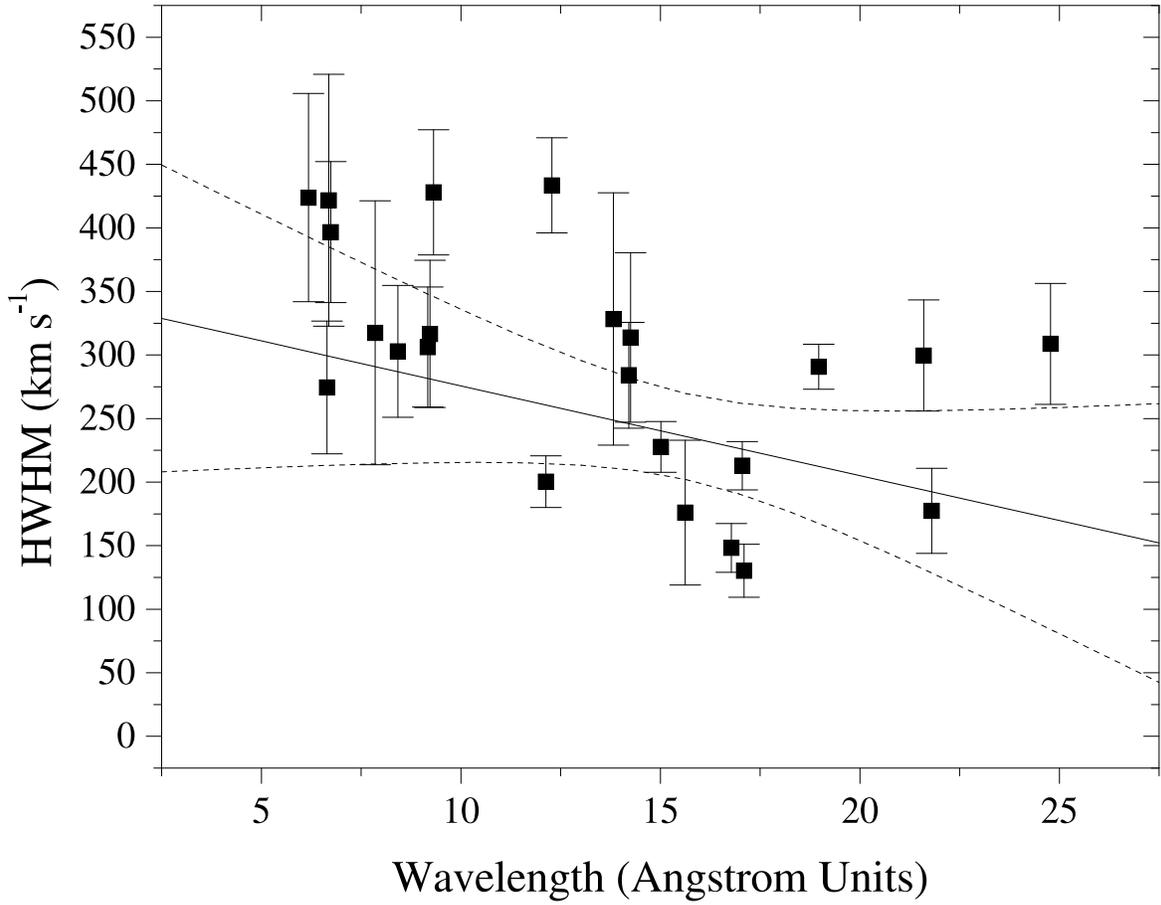}
\end{center}
\caption{HWHM for the strongest unblended lines
  in the \hetgs\/ spectrum of \tsco.  The best-fit linear function is
  indicated by the solid line with the 95 \% confidence limits shown
  as dashed lines.}  
\label{fig:hwhms}
\end{figure}

The mean HWHMs of the strongest, unblended lines are roughly 300 km
s$^{-1}$, or 15\% of the terminal velocity of the wind of \tsco\/ (see
Fig. \ref{fig:hwhms}). There is a slight trend of decreasing line
width with wavelength, which is the opposite of what is seen in the O
stars (Kahn \etal\ 2001; Waldron \& Cassinelli 2001; Cassinelli \etal\ 
2001; Miller \etal\ 2002).  In the O stars, the trend of increasing
line width with wavelength can be understood in terms either of
temperature stratification (lower temperature plasma farther out in
the wind, traveling at higher velocities) or an absorption effect (low
energies are preferentially attenuated, so these lines tend to be
broader because only photons from the outer wind, where velocities
tend to be high, can escape).  The lack of such a trend in \tsco\/
might be evidence of the lack of any wind continuum absorption effects
on the observed spectral lines.  The wind is expected to be optically
thin down to $r \lesssim 1.5$ \rstar\/ near 0.5 keV and effectively
completely optically thin at photon energies of 1 keV and above.


\subsection{Radiation-Sensitive Line Ratios}

The He-like forbidden-to-intercombination line strength ratio is known
to be sensitive to the local mean intensity of the UV radiation field
(Gabriel \& Jordan 1969; Blumenthal, Drake, \& Tucker 1972; Waldron \&
Cassinelli 2001; Kahn \etal\ 2001).  If the photospheric flux is
strong enough, then electrons can be radiatively excited out of the
metastable upper level of the forbidden line ($^3S_1$) to the upper
level of the intercombination line ($^3P_{1,2}$), weakening the
forbidden line and strengthening the intercombination line.  Thus a
measurement of the \ftoi\/ ratio [using the more standard
spectroscopic notation, $z/(x+y)$], potentially yields the radius of
the X-ray emitting plasma (via sensitivity of the mean intensity to
the dilution factor).  This assumes that we know or can ignore limb
darkening and other geometrical effects and that we know the UV
luminosity of the star at the relevant wavelengths.

More traditionally, this line ratio has been used as a density
diagnostic in coronal sources (where, because of the low temperature
of the stars, very little UV radiation is present).  Of course, in hot
stars, this line ratio is still sensitive to density (via collisional
excitation between the same two levels described above).  However, the
densities that would be required to explain the small observed \ftoi\/
values are very large, while the radiation field requirements are much
more reasonable.

In order to extract information from these line ratios, we model the
oxygen, neon, magnesium, and silicon excitation/ionization using
NLTERT, a non-LTE radiation transport and statistical equilibrium code
(described in MacFarlane \etal\ 1993; MacFarlane, Cohen, and Wang
1994).  We use a Kurucz model atmosphere, constrained by observational
data in the spacecraft UV (relevant for O and Ne), but unconstrained
by direct observation in the far-UV (Mg) and extreme-UV (EUV) (Si and
S).  We note that the model atmosphere fluxes are sensitive to the
adopted temperature only in the EUV, in which a change of 1000 K can
lead to about a factor of 2 change in emergent flux.

\begin{figure}
\begin{center}
\includegraphics[angle=90,scale=0.25]{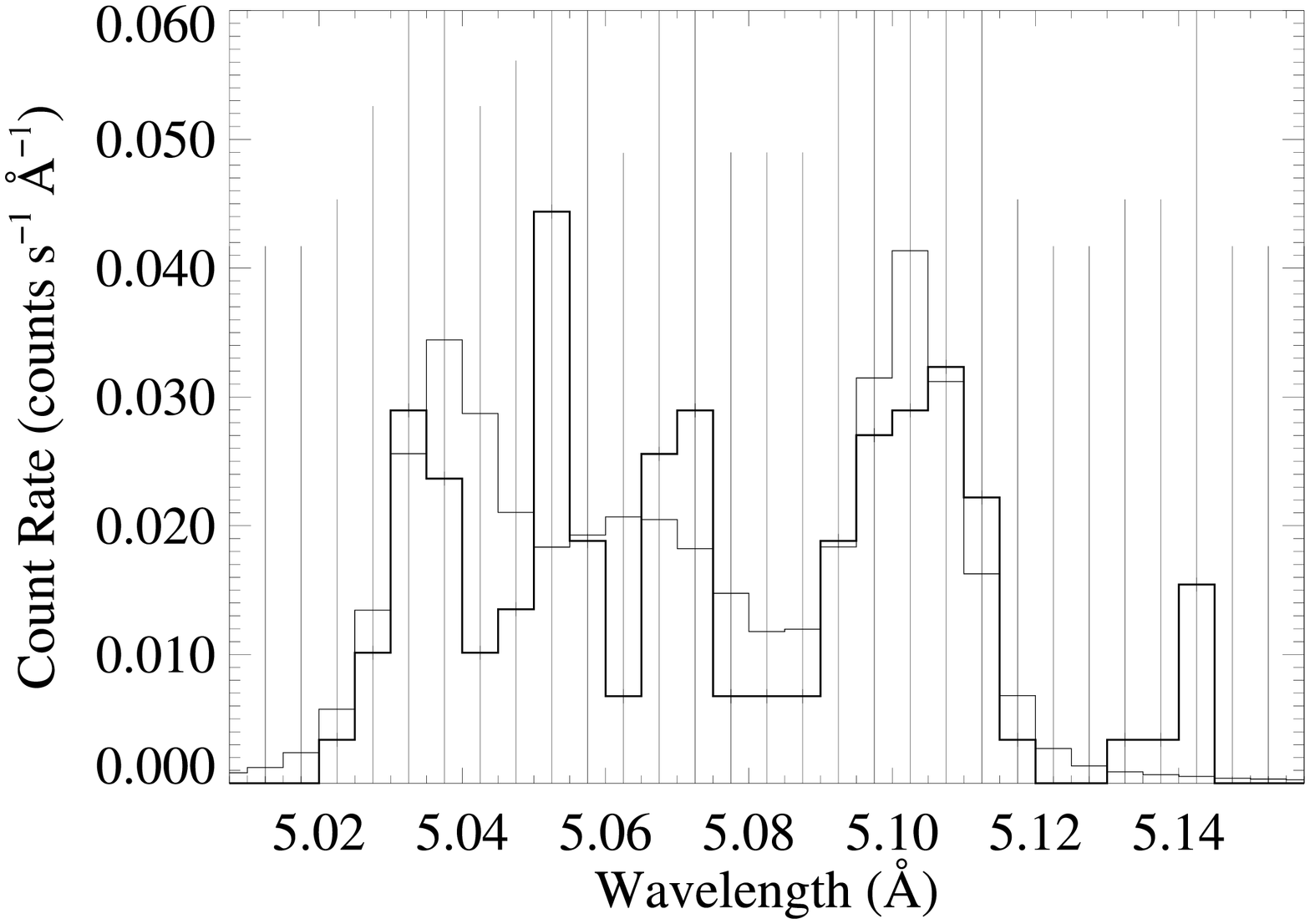}
\includegraphics[angle=90,scale=0.25]{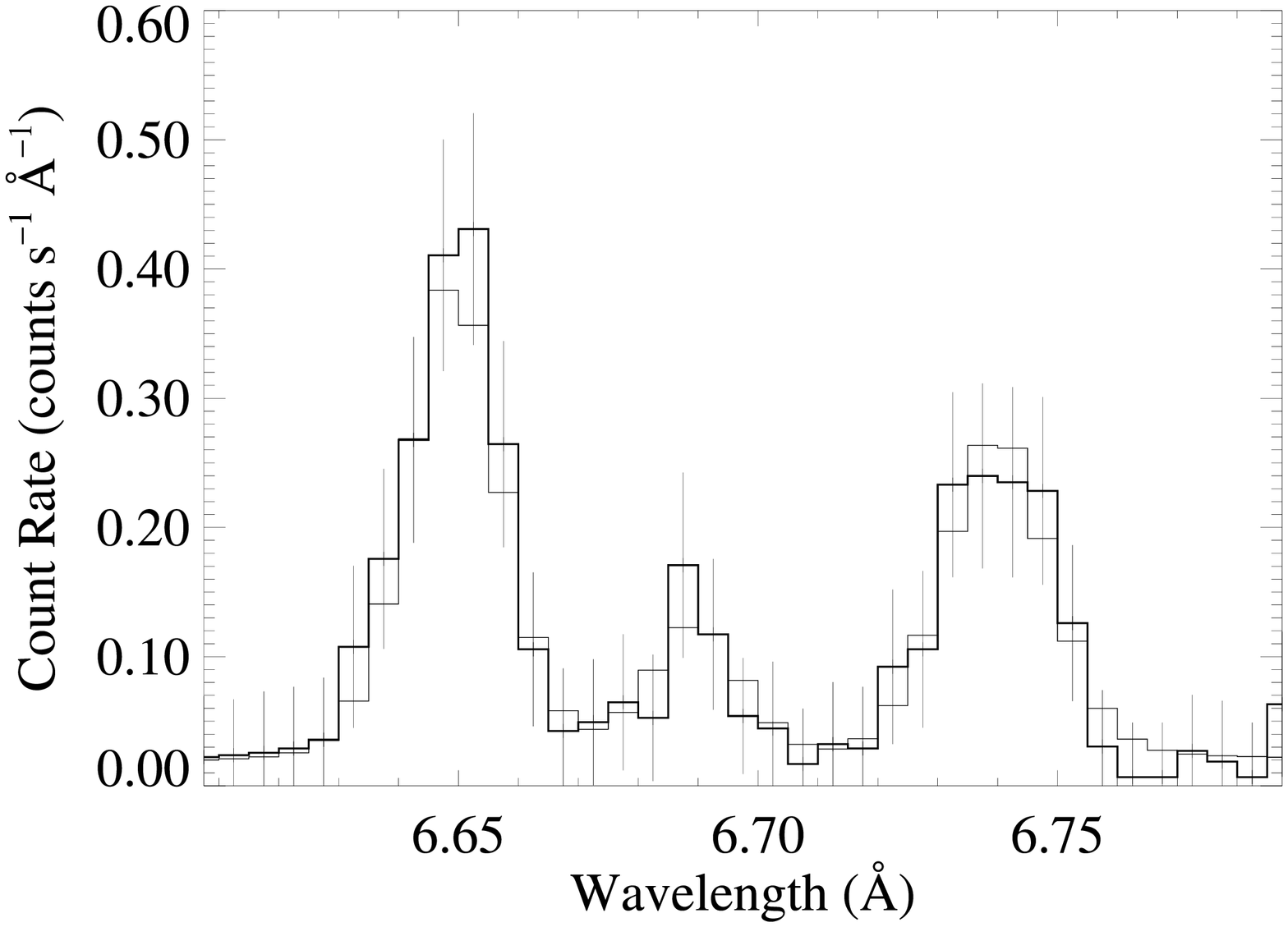}
\includegraphics[angle=90,scale=0.25]{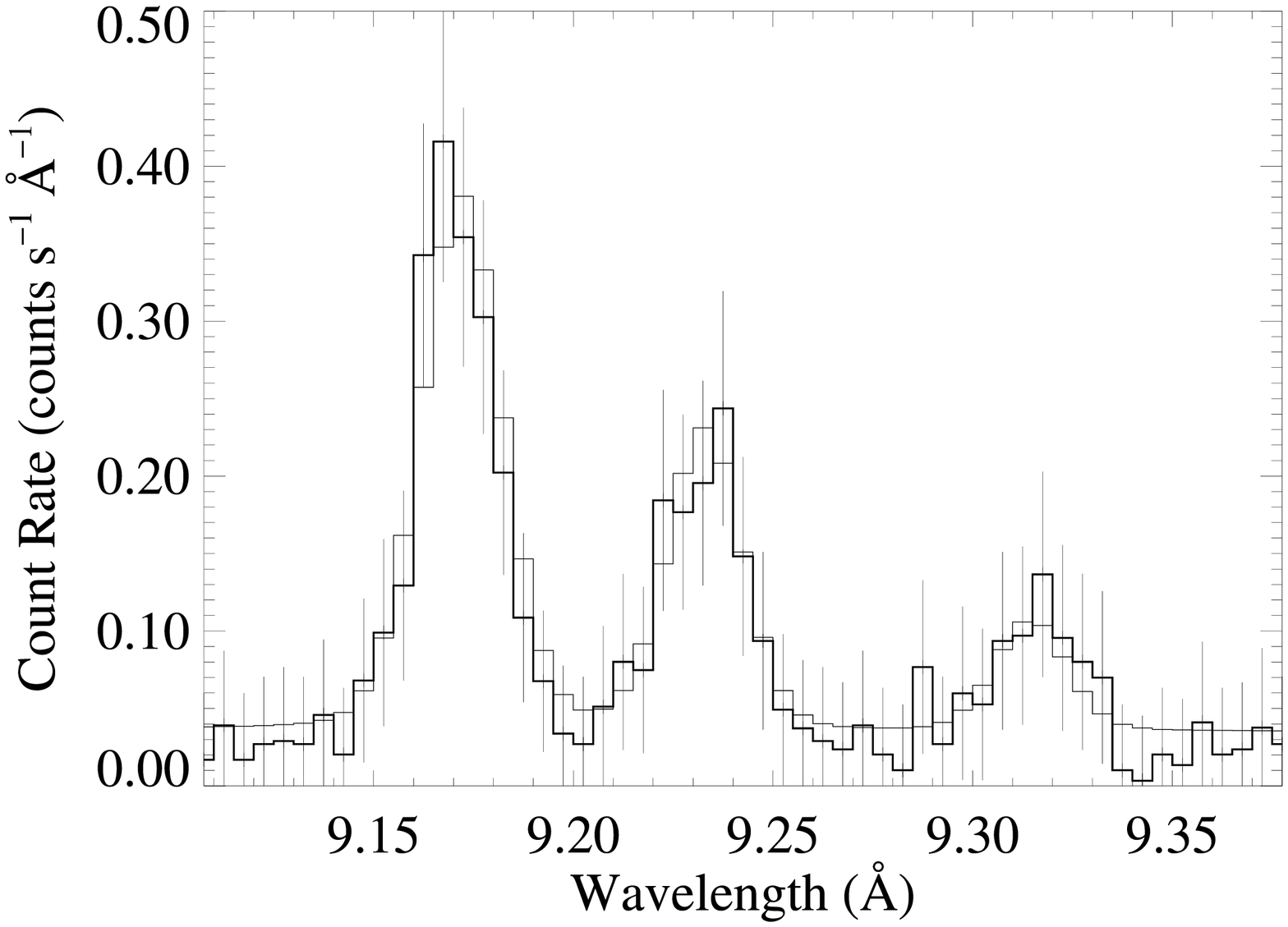}
\includegraphics[angle=90,scale=0.25]{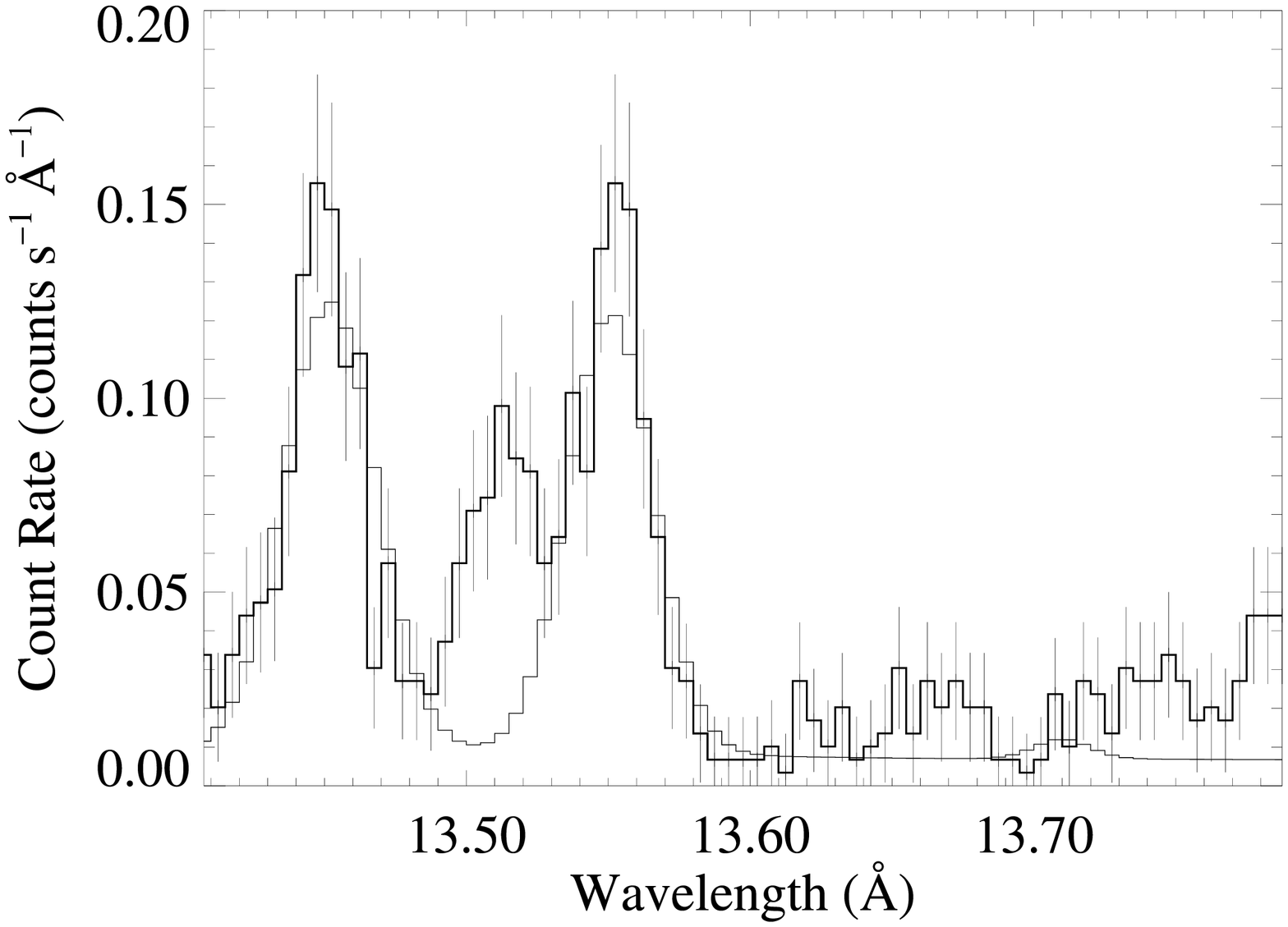}
\includegraphics[angle=90,scale=0.25]{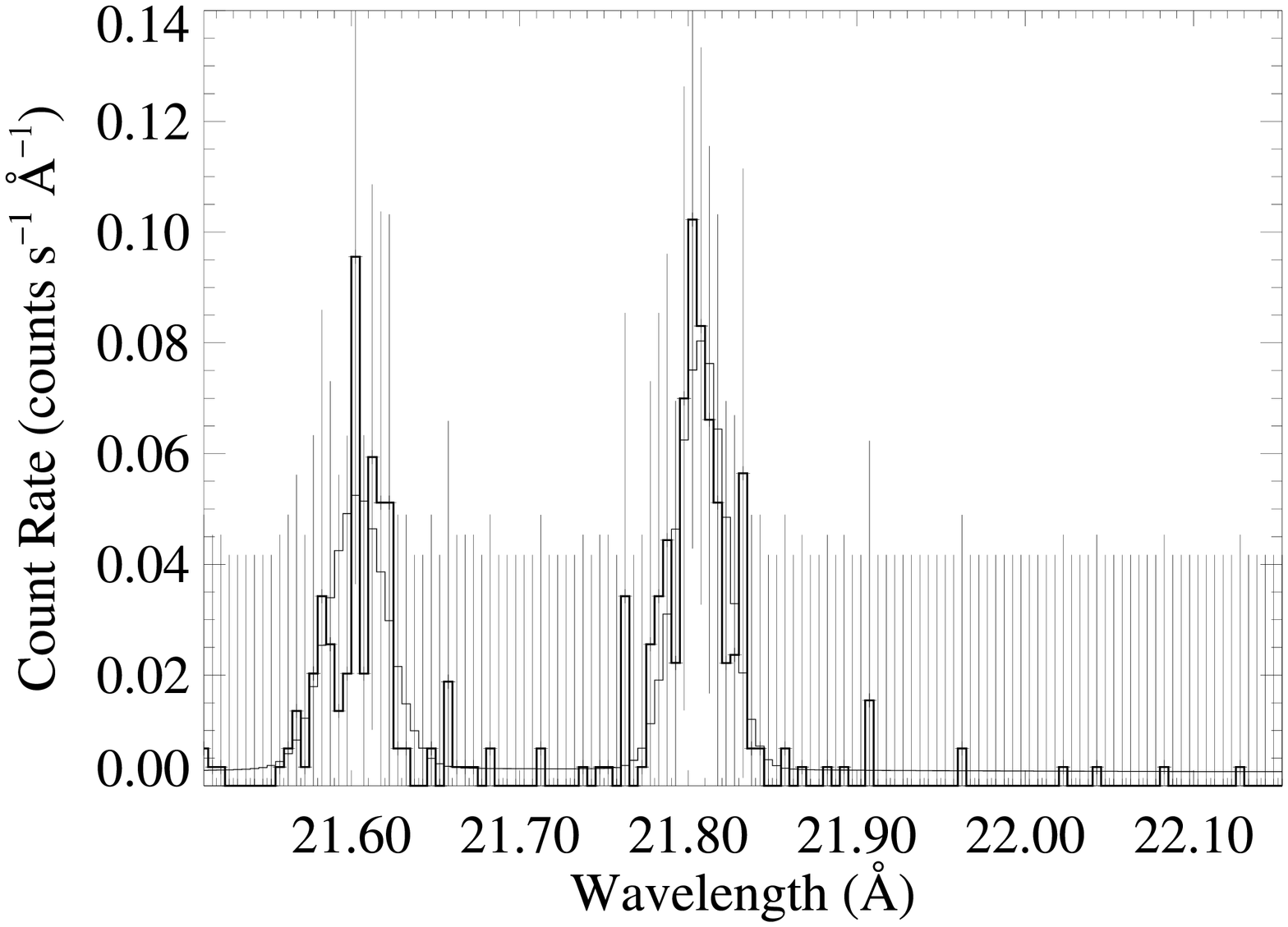}
\end{center}
\caption{MEG +1 and -1 order data for the helium complexes of
  \ion{S}{15}, \ion{Si}{13}, \ion{Mg}{11}, \ion{Ne}{9}, and \ion{O}{7}
  along with the best-fit models. Note that the error bars shown here
  (Sherpa's ``chi Gehrels'' option) represent Poisson noise plus a
  constant.  We used the Cash C statistic as our goodness-of-fit
  indicator, and this statistic does not make use of the formal
  errors.}
\label{fig:Healpha}
\end{figure}

\begin{figure}
\begin{center}
\includegraphics[angle=270,scale=0.7]{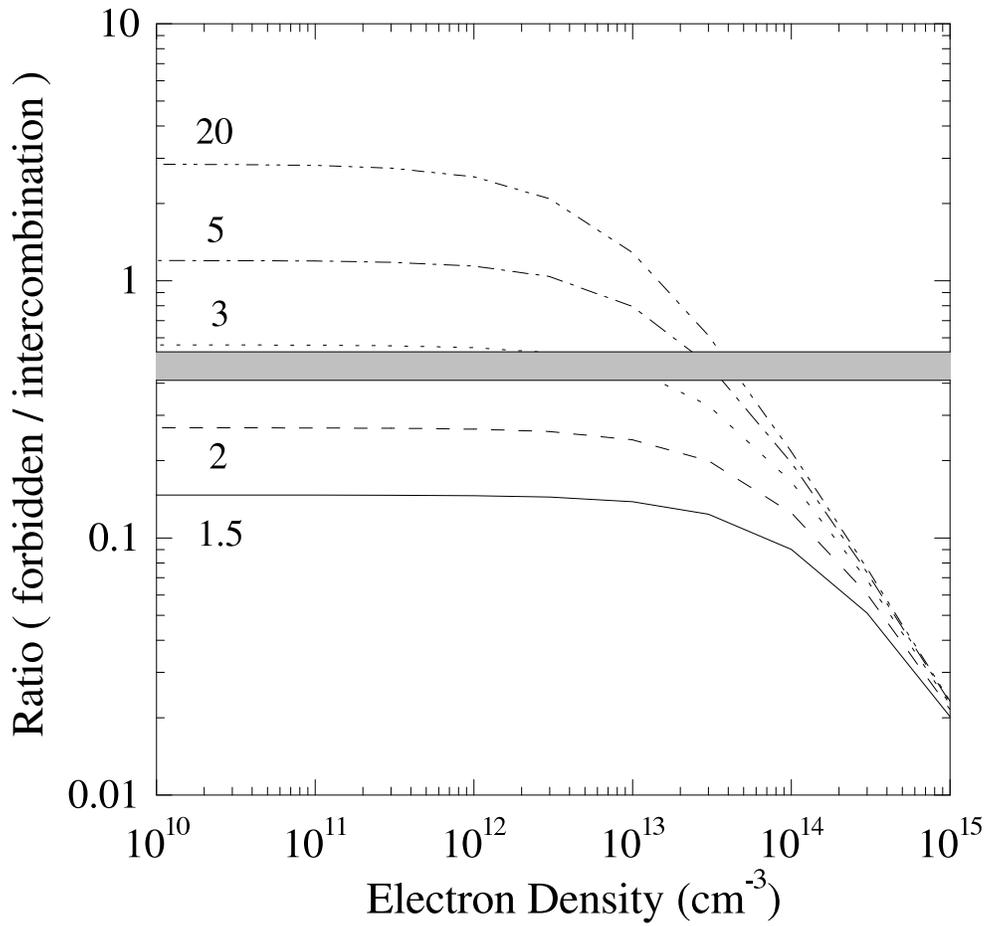}
\end{center}
\caption{Calculation of the density and mean-intensity sensitivity of
  the \ftoi\/ line ratio for \ion{Mg}{11}.  Note that the sensitivity
  to the mean intensity enters via the distance from the photosphere.
  The distance of the plasma from the photosphere is indicated for
  each model in units of the stellar radius.  The range from the data
  is indicated by the shaded area.}
\label{fig:fir_model_mg}
\end{figure}

The He$\alpha$ line complexes are shown in Figure \ref{fig:Healpha}.
When the metastable level is neither collisionally nor radiatively
destroyed (i.e., the ``low-density limit''), the forbidden lines tend
to be about a factor of 3 stronger than the intercombination lines.
We can see in the figure that the forbidden line is heavily affected
in oxygen and neon, less so in magnesium, and less still in silicon.
The \ion{S}{15} data are marginal, but are shown for the sake of
completeness.  In Figure \ref{fig:fir_model_mg} we show the results of
our model calculation for He-like magnesium.  Note that we model the
effects of both radiation and collisions (i.e., showing both the
radial sensitivity and the density sensitivity) in these calculations.
Similar calculations were carried out for the He-like features for the
other four species.  The results of these calculations and the
constraints they imply for the location of the X-ray emitting plasma
are summarized in Table \ref{tab:fir}.  The quoted errors on the
determinations of the radii of line formation include statistical
errors only.  The uncertainties in the atmosphere model and the
associated UV luminosity will certainly affect the results for
\ion{Si}{13} (and \ion{S}{15}) more than for the other three line
ratios.  For example, if the flux were actually twice as high as in
the model atmosphere, then the lower limit to the radius of line
formation would move out from $r=1.1$ \rstar\/ to almost 1.5
\rstar.

We also point out that, where other lines were present (primarily iron
lines near the neon complex), we fitted them at the same time we
fitted the He-like features, and upper limits to the nondetected
forbidden lines were determined by fitting a Gaussian model with its
centroid position fixed at the laboratory wavelength and its width
fixed at the value derived for the corresponding He-like resonance
line.




In general, it can be seen that most of these line ratio diagnostics
are consistent with the X-ray emitting plasma being within several
stellar radii of the photosphere.  The oxygen and sulfur complexes do
not really provide interesting constraints, as they are consistent
with the X-ray emitting plasma being either right above the
photosphere or far out in the wind flow.  The neon ratio implies that
the plasma giving rise to this complex is within a stellar radius of
the photosphere, but confusion and blending with nearby iron lines may
make this result somewhat less firm than the formal errors imply. The
strongest constraints are provided by the \ion{Mg}{11} and
\ion{Si}{13} features.  The \ion{Mg}{11} \ftoi\/ ratio requires the
X-ray emitting plasma be at least 1.6 \rstar\/ above the photosphere
(but no more than 1.9 \rstar).  The \ion{Si}{13} \ftoi\/ ratio
indicates that the hotter plasma giving rise to this feature is closer
to the star, a trend that is also seen in the O stars observed with
\chandra\/ (Waldron \& Cassinelli 2001; Cassinelli \etal\/ 2001;
Miller \etal\ 2002).

Finally, we note that the intercombination line fluxes exceed the
resonance line fluxes in the case of oxygen and neon.  While the
intercombination line strength is generally expected to be less than
that of the resonance line in collisional plasmas, the observed G
ratios ($(i+f)/r$) for these complexes are not inconsistent with a
collisional plasma, since the enhanced intercombination line flux
comes at the expense of the weakened forbidden line (Pradhan \& Shull
1981).

\subsection{Density-Sensitive Line Ratios}

While the helium-like forbidden-to-intercombination ratios are
sensitive to both density (via the collisional destruction of the
$^3S$ level) and the local radiation field (through photoexcitation
from the same level), we have argued in \S5.2 that in hot stars like
\tsco, the ratio is dominated by radiative effects.  This, of course,
makes them relatively useless as density diagnostics, although, for
He-like ions for which the forbidden line is strong, an upper
limit can be placed on the density of the X-ray-emitting plasma (as
well as on the local value of the mean intensity). The He-like
complexes we observe imply upper limits of $n_{\rm e} \approx 10^{13}$
- $10^{14}$ cm$^{-3}$.

It is possible to determine an upper limit for the density via the
\ion{Fe}{17} 17.10-to-17.05 \AA\/ line ratio.  The upper level of the
17.10 \AA\/ line is metastable, and so this ratio is density sensitive
with a critical density of about $n_e \approx 10^{14}$ cm$^{-3}$
(Mauche, Liedahl, \& Fournier 2001), but the energy spacing of this
deexciting transition is larger than that for the He-like
diagnostics, so that photoexcitation requires a photon with a
wavelength below 400 \AA.  Thus, for all practical purposes, in a B
star this diagnostic is {\it not} sensitive to photospheric UV
photoexcitation.  In the \hetgs\/ spectrum of \tsco, we see the
17.10-to-17.05 \AA\/ ratio in the low-density limit.  Thus there is no
evidence for very dense ($n_e \gtrsim 10^{14}$ cm$^{-3}$) X-ray
emitting plasma on \tsco.

\subsection{Line Centroids}

\begin{figure}
\begin{center}
\includegraphics[angle=270,scale=0.7]{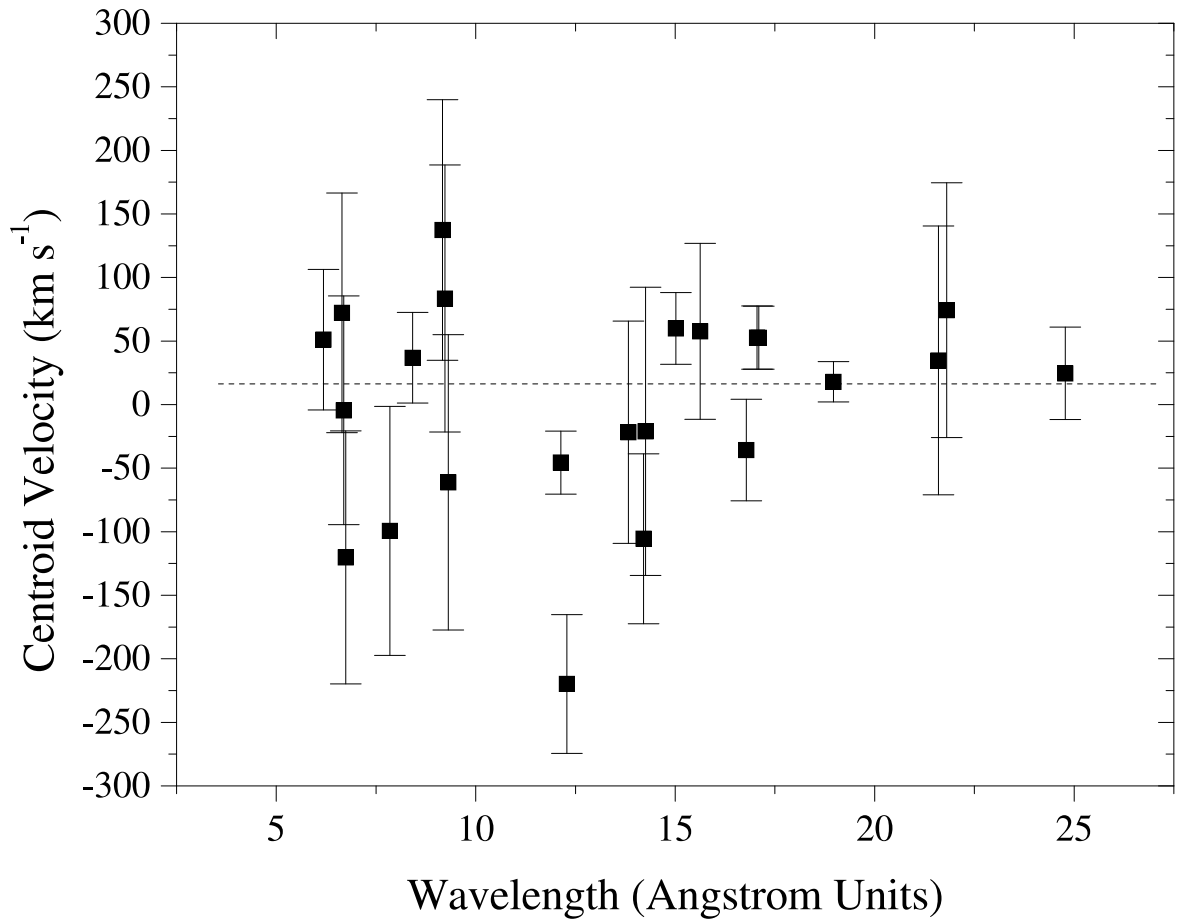}
\end{center}
\caption{Line centroids derived from fitting Gaussian models to the
  strongest lines in the \hetgs\/ spectrum. The dashed line indicates
  the weighted mean centroid velocity.}
\label{fig:centroid}
\end{figure}

In fitting the lines in the grating spectrum, we allowed the line
centroid wavelength to be a free parameter.  These results, along with
line strengths and widths, are listed in Table \ref{tab:lines}.
Although each individual line centroid is roughly consistent with the
laboratory value of the wavelength, when taken as a whole, there is
the potential to see an aggregate wavelength shift in the ensemble of
lines.  We calculated the weighted mean centroid velocity for the
roughly two dozen strongest unblended lines and found a value of
$v_{\rm centroid} = 14 \pm 61$ km s$^{-1}$ (see Fig.
\ref{fig:centroid}).  The heliocentric correction actually shifts this
number to the blue by about $-27$ km s$^{-1}$, making the heliocentric
weighted-mean line centroid velocity $-13 \pm 61$ km s$^{-1}$. Thus,
there is no direct evidence for redshifted emission that might be
expected from the shocking of infalling clumps, as in the Howk \etal\/
(2000) model, which could be associated with the wind component that
gives rise to the observed redshifted UV absorption.  We cannot,
however, rule out redshifted X-ray emission with net velocities of the
order of 100 km s$^{-1}$.


\subsection{Temperature Distribution in the Hot Plasma}

We are deferring detailed DEM modeling of the global HETG spectrum to
a future paper.  The three-temperature MEKAL (Mewe, Kaastra, \&
Liedahl 1995) collisional equilibrium model that fitted the \asca\/
spectrum of this star (Cohen \etal\ 1997b), does an adequate job of
matching the gross properties of the \hetgs\/ spectrum.  Of course,
with the vast increase in spectral resolution with \chandra, this
model does not fit in detail.  However, it is clear that a detailed,
global fit to the spectrum will not drastically alter the results from
the fit to the \asca\/ spectrum.  One question left open by the
analysis of the medium-resolution \asca\/ data is the extent to which
the high-temperature component of the spectrum is present, in terms of
both the overall emission measure of the hot component and the actual
maximum plasma temperature.

We find no obvious evidence for plasma with temperatures above the 27
MK lower limit to the hottest temperature component that was found in
the \asca\/ spectrum, and in fact, our detection of the \ion{S}{15}
feature, which is relatively prominent in the \asca\/ spectrum, is
quite marginal in the \hetgs\/ spectrum.  This is perhaps not
surprising, however, given the much lower detector effective area
available with \hetgs\ + ACIS-S compared to \asca\/ Solid-State
Imaging Spectrometer.

We have made a rough assessment of the temperature distribution
implied by the H-like to He-like line ratios from O, Ne, Mg, and Si.
We modeled these ratios using the same NLTERT non-LTE
ionization/excitation code (MacFarlane \etal\ 1993) that we used to
model the He-like {\it f/i} ratios.  The best-fit temperature for each
of these ions can be determined from the observed line ratios under
the assumption of coronal equilibrium.  We found temperatures ranging
from less than 3 MK for oxygen to slightly more than 10 MK for the
silicon.  This is in agreement with the expectations from the
off-the-shelf collisional equilibrium codes, such as APEC (Smith
\etal\ 2001), and it represents a wider temperature range than is
seen in the O stars \zpup\/ and \zori\/ based on this same diagnostic.
Of course, the presence of a wide range of plasma temperatures is not
surprising, and by itself is not a strong discriminant among the
various physical models of X-ray production in hot stars.

Finally, the wavelength region between 10 \AA\/ and 12 \AA\/ contains
several lines from high ionization stages of iron, including
\ion{Fe}{23} at 11.02 \AA\/ and 11.74 \AA\/ and and \ion{Fe}{24} at
10.62 \AA.  The equilibrium temperatures of peak emissivity of these
lines are roughly 20 MK (APED; Smith \etal\ 2001).  These lines are
quite strong in the spectrum of \tsco.  Of the other hot stars
observed thus far with \chandra\/ \hetgs\/, only \tori\/ shows these
features as strongly as does \tsco. Thus, there is evidence in the
\chandra\/ spectrum of \tsco\/ for a not-inconsequential amount of
very hot plasma, qualitatively confirming the earlier \asca\/ results,
and demonstrating a similarity between \tsco\/ and \tori\/ that sets
both of these young hot stars apart from the more standard O
supergiants.

\subsection{Time Variability}

Active coronal sources, including pre-main-sequence stars, show
frequent and strong flaring activity, often in conjunction with very
high plasma temperatures (e.g., AB Dor: Linsky \& Gagn\'{e} 2001;
pre-main-sequence stars: Montmerle \etal\ 2002). Wind-shock
sources show very little variability, and whatever variability is seen
has been interpreted as being stochastic and reflecting a Poisson
statistical description of the number of individual sites of X-ray
emission within the wind (Oskinova \etal\ 2001).  Variability levels
of 10\%, as are seen in the \chandra\/ spectra of \tsco, could
then be interpreted as an indication that there are roughly 100
individual sites of X-ray emission at any given time on this star.  Of
course, arguments about characteristic timescales (of, e.g., cooling),
as well as the evolution of the physical properties of individual
emission sites, have a bearing on the details of this interpretation.

Our detection of variability in the oxygen Ly$\alpha$ line at 18.97
\AA, but in no other line, could indicate that the relatively cooler
gas that gives rise to this feature has a more variable emission
measure than do other lines, and thus \ion{O}{8} might be present in
fewer discrete sites within the X-ray emitting region, whether it is
wind shocks or magnetically confined coronal plasma. It could, on the
other hand, be an indication that the level of absorption by the
overlying cool wind is variable.  The long-wavelength lines like
\ion{O}{8} are the most likely lines to be subject to wind absorption.
However, the expectation is that there is very little continuum
absorption of the X-rays in a hot star like \tsco, which has a
relatively low-density wind.  We note that there is enough signal in
several other lines (\ion{Si}{13}, \ion{Mg}{11}, several \ion{Fe}{17}
lines, for example) that variability at the level seen in the
\ion{O}{8} line would be detected if it were present.  We also note
that the time dependence of the variability in this line tracks the
overall X-ray variability in the spectrum as a whole and also that
the even longer wavelength lines of \ion{O}{7} and \ion{N}{7} do not
have enough signal for any variability to be detected.

\section{Discussion}

The qualitative impression one gets from comparing the \hetgs\/
spectrum of \tsco\/ to that of other stars is that this early B star
has more in common with late-type coronal X-ray sources than with
early-type wind sources.  This impression is borne out primarily in
the analysis of the line widths.  The emission lines are indeed
narrow, although they are not as narrow as the same lines seen in
Capella and AB Dor.  The bulk flow velocities derived from these
lines, while supersonic, are only 10\% - 20\% of the wind terminal
velocity.

If embedded in a standard CAK wind, the X-ray-emitting regions would
have to be within just a few tenths of a stellar radius from the
photosphere based on the observed line widths.  However, the
He-like $f/i$ ratios put constraints on the proximity of the X-ray
emitting plasma to the photosphere.  These results are consistent with
a height of 1\rstar-2\rstar\/ ($2 \rstar < r < 3 \rstar$), and for
\ion{Mg}{11} the plasma must be at least 1.6 \rstar\/ above the
photosphere.  There is thus a contradiction, that would seem to imply
the presence of hot plasma out to a few stellar radii above the
surface of \tsco, but moving with very small (but nonzero)
line-of-sight velocity.  This picture is inconsistent with standard
wind-shock scenarios (which seem to apply to O supergiants like \zpup)
in which hot plasma is embedded within the massive line-driven wind.
Our data show lines far too narrow for this to be the case.  A similar
situation is seen in the O9.5 star $\delta$ Ori, though the emission
lines in that star are significantly broader than what is seen in the
spectrum of \tsco\/ (Miller \etal\ 2002).

It is also difficult, however, to reconcile these data with a generic,
solar-type picture of magnetic X-ray activity, in which small magnetic
loops confine plasma heated by magnetic reconnection and the
deposition of mechanical energy.  In this picture, the plasma would be
closer to the UV-bright photosphere of \tsco\/ than can be accounted
for by the observed $f/i$ value for magnesium.  Furthermore, we see no
evidence for coronal X-ray flaring in our data.  It is conceivable
that relatively stable, large scale magnetic loops could exist on the
surface of \tsco, with hot plasma confined at the tops of the loops,
several stellar radii from the surface, but the standard picture of
energetic, relatively small spatial scale magnetic activity, as seen
in the sun, is not consistent with these observations of \tsco.

One scenario that could explain the two primary results from the
analysis of the \chandra\/ \hetgs\/ spectrum of \tsco\/ is the ``clump
infall'' model (Howk \etal\/ 2000), which was put forward to explain
the hard \asca\/ spectrum and the redshifted UV absorption features
seen in this star.  In this model it is {\it assumed} that density
enhancements (``clumps'') form in the wind of a hot star.  The authors
explored the dynamics of such clumps under the combined influence of
radiation driving, gravity, and wind drag.  They found that for the
parameters appropriate to \tsco, such density enhancements stall at
heights of several stellar radii, and fall back toward the star.  A
bow shock would develop around each clump, leading to strong shock
heating and presumably hard X-ray emission.  The infalling clumps then
also explain the observed redshifted UV absorption features.  Since
the outflowing wind, when it passes through the shock front associated
with one of these density enhancements, would be slowed by a factor of
4, this scenario provides a means for explaining the lack of strong
line-broadening seen in our \chandra\/ spectrum, and the heights at
which these clumps stall are consistent with the He-like \ftoi\/
ratios.  It should be stressed, however, that Howk \etal\/ (2000)
present no specific mechanism for the condensation of these density
enhancements out of the wind.

The relative youth of \tsco\/ (Kilian 1994) has suggested that a
fossil magnetic field may be present in this star.  The very low
projected rotational velocity is also consistent with magnetic
spin-down (or, of course, with a pole-on viewing angle). Recent
modeling of radiation-driven winds in the presence of a dipole
magnetic field (Babel \& Montmerle 1997; Donati \etal\ 2002; ud-Doula
\& Owocki 2002) has suggested another scenario that may also be
relevant for \tsco\/ and could explain the formation of density
enhancements at several stellar radii in this star's wind.  The
magnetically confined wind shock (MCWS) model predicts the confinement
of the stellar wind by a strong enough dipole field and the subsequent
shock heating of wind material trapped in the field and forced to
collide at the magnetic equator.  In the case of \tsco, a field of
less than 100 G is required to provide the necessary confinement
(ud-Doula \& Owocki 2002).  This model has been successfully applied
to \tori\/ (Donati \etal\ 2002), which is a very young O star with
X-ray properties similar to those of \tsco.

Dynamic simulations have shown that plasma in the post-shock region in
the MCWS model can eventually fall back toward the star when it can no
longer be supported by radiation pressure and magnetic tension
(ud-Doula \& Owocki 2002).  By coupling this mechanism to the Howk
\etal\ (2000) scenario, we may have a self-consistent, physical
picture of the strong shock heating of wind plasma, in a nearly
stationary equatorial zone, at a height of several stellar radii, with
periodic infalling of material.  We await dynamical simulations of
this particular star in order to make quantitative comparisons with
our observational data.  In addition, synthesis of X-ray line profiles
from the dynamical simulations would be very important, since it is
not clear if the specific observed small line widths can be produced
within this context.

The other properties we derived from the \tsco\/ \hetgs\/ spectrum can
perhaps also be explained in this scenario of magnetic wind shock
heating and confinement combined with the infall of density
enhancements in the wind. The high plasma temperatures are naturally
explained by the large shock velocities inherent in the collision
between the fast stellar wind and the equatorial shocked disk (as well
as with the infalling clumps).

The lack of very dense plasma argues against very dense coronal-type
magnetic structures near the photosphere, although some level and type
of coronal heating mechanism is certainly not ruled out.  In fact,
recent work (Charbonneau \& MacGregor 2001; MacGregor \& Cassinelli
2002) suggests that a dynamo mechanism could be operating in the
interiors of normal OB stars and that if a means of bringing the
dynamo-generated magnetic fields to the surface could be identified,
then solar-type coronal activity on hot stars may be quite plausible.
The modest time variability seen in the \chandra\/ spectrum is more
significant than that seen in O stars that are supposed to be wind
X-ray sources.  This variability level (rms variations at the 10\%
level) may simply be indicative of a relatively small number of sites
of X-ray production on \tsco\/ as compared with O stars showing a more
standard mode of wind-shock X-ray production (Oskinova \etal\ 2001),
and no large flare events were seen in this observation of \tsco\/ (or
in any of the earlier X-ray observations of this star).  Therefore,
any coronal model applied to \tsco\/ would need to explain the high
temperatures (presumably with a strong dynamo mechanism of some sort)
but also the absence of frequent strong flaring.  In addition, such a
coronal mechanism would need to explain the presence of bulk plasma
velocities of several hundred km s$^{-1}$, along with at least some
X-ray-emitting plasma a significant height above the surface of the
star.

In summary, the unusual X-ray properties of \tsco, elucidated in great
detail for the first time by the superior spectral resolution of the
\chandra\/ \hetgs, now require detailed dynamical modeling in order to
test the viability of the MCWS model, deduce the role of clump infall,
and test the alternate scenario of dynamo activity.  However, we can
say with certainty that the standard picture of numerous shocks
embedded in a fast stellar wind cannot explain the narrow X-ray lines
and the formation heights of several stellar radii of the very hot
X-ray emitting plasma on this young hot star.

\acknowledgements

We wish to thank Eric Levy, Carolin Cardamone, James Ciccarelli, and
Prudence Schran for help with the data analysis and Marc Gagn\'{e} for
fruitful science discussions.  This research was supported in part by
NASA grant GO0-1089A to Swarthmore College, Prism Computational
Sciences, and the University of Wisconsin. Work was performed
under the auspices of the US Department of Energy by the University
of California Lawrence Livermore National Laboratory under contract
No.\ W-7405-Eng-48.

\newpage

\newpage

\begin{deluxetable}{lcccccc}
\tablewidth{0pt}       
\tablecaption{Emission Lines in the \hetgs\/ Spectrum}
\tabletypesize
\scriptsize

\tablehead{ 
\colhead{Ion} & 
\colhead{$\lambda_{lab}$} &
\colhead{$\lambda_{obs}$} & 
\colhead{Flux} &
\colhead{Half-Width\tablenotemark{a}} & 
\colhead{Centroid Velocity\tablenotemark{b}} & 
\colhead{log $T_{\rm max-emiss}$\tablenotemark{c}} 
  \\
\colhead{} & 
\colhead{(\AA)} & 
\colhead{(\AA)} & 
\colhead{($10^{-5}$ ph s$^{-1}$ cm$^{-2}$)} & 
\colhead{(km s$^{-1}$)} &
\colhead{(km s$^{-1}$)} &
\colhead{(K)} } 

\startdata



\ion{S}{15}\tablenotemark{d} & $5.0387 \pm 0.002$ & \nodata  &
 $0.936^{+.270}_{-.235}$ & \nodata
 & \nodata & 7.1  \\

\ion{S}{15}\tablenotemark{d} & $5.0631 \pm 0.002$ & \nodata &
 $0.464^{+.234}_{-.196}$ & \nodata & \nodata & 7.1  \\

\ion{S}{15}\tablenotemark{d} & $5.1015 \pm 0.002$ & \nodata &
 $0.944^{+.268}_{-.234}$ & \nodata & \nodata & 7.1  \\

\ion{Si}{14} & $6.1821 \pm 0.0003$ & $6.1832 \pm 0.0011$ & $2.267^{+.199}_{-.188}$ &
$424^{+84}_{-80}$ & $51^{+53+15}_{-53-15}$ & 7.2  \\

\ion{Si}{13} & $6.6479 \pm 0.002$ & $6.6495 \pm 0.0006$ &
$3.771^{+.220}_{-.211}$ & $275^{+51}_{-54}$ & $72^{+27+90}_{-27-90}$ &
7.0 \\

\ion{Si}{13} & $6.6882 \pm 0.002$ & $6.6881 \pm 0.0014$ &
$1.439^{+.143}_{-.134}$ & $422^{+106}_{-93}$ & $-4^{+63+90}_{-63-90}$ &
7.0 \\

\ion{Si}{13} & $6.7403 \pm 0.002$ & $6.7376 \pm 0.001$ &
$3.014^{+.236}_{-.224}$ & $397^{+57}_{-54}$ & $-120^{+45+89}_{-45-89}$ &
7.0 \\

\ion{Fe}{18} & $7.0889 \pm n/a$ & $7.0952 \pm 0.006$ &
$0.378^{+.159}_{-.067}$ & $169^{+211}_{-169}$ & $266^{+254}_{-72}$ &
7.2 \\

\ion{Mg}{12} & $7.1062 \pm 0.00004$ & $7.1107^{+0.0022}_{-0.0032}$ &
$0.446^{+.137}_{-.058}$ & $410^{+148}_{-190}$ & $192^{+93+2}_{-135-2}$ &
7.0 \\

\ion{Mg}{11} & $7.8503 \pm 0.002$ & $7.8477 \pm 0.0016$ &
$0.827^{+.110}_{-.101}$ & $317^{+106}_{-101}$ & $-99^{+61+76}_{-61-76}$ &
6.8 \\

\ion{Al}{12} & $7.8721 \pm 0.003$ & $7.8814 \pm 0.0032$ &
$0.339^{+.085}_{-.074}$ & $189^{+201}_{-189}$ & $354^{+114+122}_{-114-122}$ &
6.9 \\

\ion{Mg}{12} & $8.4210 \pm 0.0001$ & $8.422 \pm 0.001$ &
$2.802^{+.340}_{-.104}$ & $303^{+52}_{-52}$ & $37^{+36+2}_{-36-2}$ &
7.0 \\

\ion{Mg}{11} & $9.1688 \pm 0.003$ & $9.173 \pm 0.001$ &
$4.230^{+.289}_{-.278}$ & $306^{+49}_{-46}$ & $137^{+29+98}_{-29-98}$ &
6.8 \\

\ion{Mg}{11} & $9.2308 \pm 0.003$ & $9.2334 \pm 0.0012$ &
$2.634^{+.229}_{-.218}$ & $317^{+59}_{-57}$ & $83^{+39+98}_{-39-98}$ &
6.8 \\

\ion{Mg}{11} & $9.3143 \pm 0.003$ & $9.3124 \pm 0.002$ &
$1.265^{+.153}_{-.145}$ & $428^{+80}_{-90}$ & $-61^{+64+97}_{-64-97}$ &
6.8 \\

\ion{Fe}{21} & $9.4797 \pm n/a$ & $9.4762 \pm 0.0014$ &
$1.306^{+.149}_{-.140}$ & $263^{+91}_{-86}$ & $-111^{+44}_{-44}$ &
7.0 \\

\ion{Ne}{10} & $10.2388 \pm 0.00005$ & $10.239 \pm 0.002$ &
$2.069^{+.239}_{-.226}$ & $294^{+67}_{-65}$ & $-5^{+59+1}_{-59-1}$ &
6.8 \\

\ion{Fe}{18} & $10.3603 \pm n/a$ & $10.359 \pm 0.002$ &
$1.151^{+.242}_{-.214}$ & $211^{+123}_{-171}$ & $-94^{+58}_{-58}$ &
6.9 \\

\ion{Fe}{18} & $10.537 \pm n/a$ & $10.536^{+0.003}_{-0.002}$ &
$0.815^{+.234}_{-.204}$ & $39^{+172}_{-39}$ & $-29^{+85}_{-57}$ &
6.9 \\

\ion{Fe}{24} & $10.619 \pm 0.02$ & $10.625 \pm 0.004$ &
$2.057^{+.367}_{-.336}$ & $482^{+151}_{-116}$ & $170^{+113+565}_{-113-565}$ &
7.3 \\

\ion{Fe}{17} & $10.77 \pm 0.002$ & $10.773 \pm 0.002$ &
$1.850^{+.343}_{-.310}$ & $192^{+84}_{-108}$ & $84^{+57+56}_{-57-56}$ &
6.8 \\

\ion{Fe}{17} & $11.254 \pm 0.002$ & $11.251 \pm 0.001$ &
$2.754^{+.287}_{-.269}$ & $273^{+59}_{-54}$ & $-80^{+27+53}_{-27-53}$ &
6.8 \\

\ion{Fe}{18} & $11.326 \pm 0.004$ & $11.321 \pm 0.002$ &
$2.975^{+.300}_{-.281}$ & $467^{+63}_{-55}$ & $-132^{+53+106}_{-53-106}$ &
6.9 \\

\ion{Ne}{9} & $11.544 \pm 0.01$ & $11.546 \pm 0.004$ &
$3.181^{+.519}_{-.477}$ & $509^{+149}_{-132}$ & $52^{+104+260}_{-104-260}$ &
6.6 \\

\ion{Fe}{23} & $11.736 \pm 0.004$ & $11.745 \pm 0.002$ &
$2.939^{+.514}_{-.470}$ & $232^{+115}_{-121}$ & $230^{+51+102}_{-51-102}$ &
7.2 \\

\ion{Fe}{22} & $11.770 \pm 0.003$ & $11.775^{+0.003}_{-0.002}$ &
$1.696^{+.419}_{-.374}$ & $67^{+175}_{-67}$ & $127^{+76+76}_{-51-76}$ &
7.1 \\

\ion{Ne}{10} & $12.134 \pm 0.0001$ & $12.132 \pm 0.001$ &
$16.832^{+.766}_{-.746}$ & $200^{+20}_{-20}$ & $-46^{+25+1}_{-25-1}$ &
6.8 \\

\ion{Fe}{21} & $12.284 \pm 0.002$ & $12.275 \pm 0.001$ &
$9.114^{+.617}_{-.593}$ & $433^{+39}_{-36}$ & $-220^{+49+24}_{-49-24}$ &
7.0 \\

\ion{Fe}{21} & $12.393 \pm 0.005$ & $12.401 \pm 0.004$ &
$1.122^{+.317}_{-.287}$ & $198^{+184}_{-164}$ & $194^{+97+121}_{-97-121}$ &
7.0 \\

\ion{Fe}{20} & $12.823 \pm 0.005$ & $12.833 \pm 0.002$ &
$7.847^{+.640}_{-.611}$ & $505^{+54}_{-47}$ & $242^{+47+117}_{-47-117}$ &
7.0 \\

\ion{Ne}{9} & $13.447 \pm 0.004$ & $13.450 \pm 0.001$ &
$10.085^{+.850}_{-.816}$ & $178^{+39}_{-41}$ & $60^{+22+89}_{-22-89}$ &
6.6 \\

\ion{Fe}{19} & $13.518 \pm 0.002$ & $13.514 \pm 0.002$ &
$8.261^{+.795}_{-.756}$ & $312^{+53}_{-47}$ & $-89^{+44+44}_{-44-44}$ &
6.9 \\

\ion{Ne}{9} & $13.553 \pm 0.005$ & $13.553 \pm 0.001$ &
$10.850^{+.860}_{-.823}$ & $157^{+36}_{-35}$ & $0^{+22+111}_{-22-111}$ &
6.6 \\

\ion{Fe}{19} & $13.795 \pm 0.005$ & $13.782 \pm 0.002$ &
$2.394^{+.502}_{-.463}$ & $114^{+115}_{-114}$ & $-283^{+44+109}_{-44-109}$ &
6.9 \\

\ion{Fe}{17} & $13.825 \pm 0.002$ & $13.824^{+0.004}_{-0.003}$ &
$3.496^{+.616}_{-.573}$ & $328^{+111}_{-88}$ & $-22^{+87+43}_{-65-43}$ &
6.8 \\

\ion{Fe}{18} & $14.208 \pm 0.003$ & $14.203 \pm 0.001$ &
$10.455^{+.884}_{-.841}$ & $284^{+42}_{-41}$ & $-106^{+21+63}_{-21-63}$ &
6.9 \\

\ion{Fe}{18} & $14.256 \pm 0.005$ & $14.255 \pm 0.002$ &
$4.681^{+.638}_{-.590}$ & $314^{+73}_{-60}$ & $-21^{+42+105}_{-42-105}$ &
6.9 \\

\ion{Fe}{18} & $14.534 \pm 0.003$ & $14.536 \pm 0.002$ &
$5.485^{+.915}_{-.824}$ & $204^{+65}_{-59}$ & $41^{+41+62}_{-41-62}$ &
6.9 \\

\ion{Fe}{18} & $14.571 \pm 0.011$ & $14.575 \pm 0.004$ &
$1.215^{+.486}_{-.390}$ & $29^{+233}_{-29}$ & $82^{+82+226}_{-82-226}$ &
6.9 \\

\ion{Fe}{17} & $15.014 \pm 0.001$ & $15.017 \pm 0.001$ &
$32.362^{+1.672}_{-1.621}$ & $228^{+20}_{-20}$ & $60^{+20+20}_{-20-20}$ &
6.7 \\

\ion{Fe}{17} & $15.261 \pm 0.002$ & $15.264 \pm 0.001$ &
$13.100^{+1.010}_{-.961}$ & $226^{+36}_{-34}$ & $59^{+20+39}_{-20-39}$ &
6.7 \\

\ion{Fe}{18} & $15.625 \pm 0.003$ & $15.628 \pm 0.002$ &
$5.151^{+.700}_{-.651}$ & $176^{+58}_{-56}$ & $58^{+38+58}_{-38-58}$ &
6.7 \\

\ion{O}{8} & $16.006 \pm 0.0001$ & $16.006 \pm 0.001$ &
$13.479^{+1.004}_{-0.957}$ & $149^{+27}_{-27}$ & $13^{+19+37}_{-19-37}$ &
6.5 \\

\ion{Fe}{18} & $16.071 \pm 0.003$ & $16.075 \pm 0.001$ &
$8.035^{+.813}_{-.766}$ & $129^{+39}_{-41}$ & $75^{+19+56}_{-19-56}$ &
6.8 \\

\ion{Fe}{17} & $16.780 \pm 0.002$ & $16.778 \pm 0.001$ &
$23.371^{+1.358}_{-1.309}$ & $148^{+19}_{-19}$ & $-36^{+18+36}_{-18-36}$ &
6.7 \\

\ion{Fe}{17} & $17.051 \pm 0.001$ & $17.054 \pm 0.001$ &
$29.207^{+1.605}_{-1.554}$ & $213^{+19}_{-19}$ & $53^{+18+18}_{-18-18}$ &
6.7 \\

\ion{Fe}{17} & $17.096 \pm 0.001$ & $17.099 \pm 0.001$ &
$26.114^{+1.534}_{-1.479}$ & $130^{+21}_{-21}$ & $53^{+18+18}_{-18-18}$ &
6.7 \\

\ion{O}{8} & $18.969 \pm 0.0001$ & $18.970 \pm 0.001$ &
$50.569^{+2.629}_{-2.541}$ & $291^{+18}_{-17}$ & $18^{+16+1}_{-16-1}$ &
6.5 \\

\ion{O}{7} & $21.602 \pm 0.007$ & $21.604 \pm 0.003$ &
$21.712^{+2.708}_{-2.506}$ & $300^{+47}_{-41}$ & $35^{+42+97}_{-42-97}$ &
6.3 \\

\ion{O}{7} & $21.804 \pm 0.007$ & $21.809 \pm 0.002$ &
$27.271^{+3.110}_{-2.896}$ & $177^{+35}_{-32}$ & $74^{+28+96}_{-28-96}$ &
6.3 \\

\ion{N}{7} & $24.781 \pm 0.0001$ & $24.783 \pm 0.003$ &
$22.019^{+3.020}_{-2.772}$ & $309^{+53}_{-42}$ & $25^{+36+1}_{-36-1}$ &
6.3 \\

\enddata

\tablenotetext{a}{Derived HWHM.}
\tablenotetext{b}{The first quoted uncertainties are statistical and
  based on the fit of the Gaussian model, while the second quoted
  uncertainties are systematic and are based on the uncertainties in
  the laboratory rest wavelengths as compiled in APED.}
\tablenotetext{c}{Temperature of maximum emissivity; from APED.}
\tablenotetext{d}{Because the position and width of the Gaussian
  models were fixed for these lines, we do not list best-fit values or
  uncertainties for these quantities.}
\label{tab:lines} 
\end{deluxetable}

\begin{table}[tb]       
\caption{Emission-Line Blends}
\vspace{0.4cm}
\begin{center}
\begin{tabular}{lcc} 
\tableline
Ion & $\lambda_{lab}$\tablenotemark{a} & Flux 
\\
& (\AA) &  ($10^{-5}$ photons s$^{-1}$ cm$^{-2}$) \\
\tableline

\ion{Fe}{20} & $10.00$ & $3.638^{+0.285}_{-0.263}$  \\

\ion{Ni}{19} & $10.58$ & $1.899^{+0.386}_{-0.342}$  \\

\ion{Fe}{18}, \ion{Fe}{19}, \ion{Fe}{24}  & $10.65$ & $0.458^{+0.212}_{-0.179}$  \\

\ion{Fe}{23} & $11.00$ & $2.367^{+0.384}_{-0.351}$  \\

\ion{Fe}{21} & $12.42$ & $4.555^{+0.463}_{-0.436}$  \\

\tableline 
\end{tabular} 
\tablenotetext{a}{Approximate mean wavelength of the blended feature.}
\end{center}
\label{tab:blends} 
\end{table}

\begin{table}[tb]       
\caption{Helium-like Forbidden-to-Intercombination Ratios}
\vspace{0.4cm}
\begin{center}
\begin{tabular}{cccc} 
\tableline
Ion & $\lambda_{UV}$\tablenotemark{a} & $f/i$ & Formation Radius\\
& (\AA) &  & (\rstar) \\
\tableline

\ion{O}{7} & 1623 & $<0.13$ & $<10$ \\

\ion{Ne}{9} & 1263 & $<0.03$ & $<2$ \\

\ion{Mg}{11} & 1036 & $0.48 \pm 0.07$ & $2.6 - 2.9$ \\

\ion{Si}{13} & 865 & $2.10^{+.28}_{-.26}$ & $1.1 - 1.5$ \\

\ion{S}{15} & 743 & $2.03^{+1.03}_{-1.14}$ & $<20$ \\

\tableline 
\end{tabular} 
\tablenotetext{a}{Wavelength of the $^3S - ^3P$ transition that
  depletes the upper level of the forbidden line and enhances the
  intercombination line.}
\end{center}
\label{tab:fir} 
\end{table}


\begin{thebibliography}{}


\bibitem[Abbott 1982]{a82} Abbott, D. C. 1982, \apj, 259, 282

\bibitem[Babel \& Montmerle 1997]{bm97} Babel, J. \& Montmerle, T. 1997, \apjl, 485, L29
  
\bibitem[Bergh\"ofer \& Schmitt 1994]{bs94} Bergh\"ofer, T. W., \&
  Schmitt, J. H. M. M.  1994, \aap, 292, L5
  
\bibitem[Blumenthal, Drake, \& Tucker 1972]{bdt72} Blumenthal, G. R.,
  Drake, G. W. F., \& Tucker, W. H. 1972, \apj, 172, 205
  
\bibitem[Cassinelli \etal\ 2002]{cbmmt02} Cassinelli, J. P., Brown, J.
  C., Maheswaran, M., Miller, N. A., \& Telfer, D. C. 2002, \apj, 578, 951
  
\bibitem[Cassinelli et al. 2001]{cmwmc01} Cassinelli, J. P., Miller,
  N. A., Waldron, W. L., MacFarlane, J. J., \& Cohen, D. H. 2001,
  \apj, 554, L55
  
\bibitem[Cassinelli \& Olson 1979]{co79} Cassinelli, J. P., \& Olson,
  G. L. 1979, \apj, 229, 304
  
\bibitem[Castor, Abbott, \& Klein 1975]{cak75} Castor, J. I., Abbott,
  D. C., \& Klein, R. I. 1975, \apj, 195, 157 (CAK)
  
\bibitem[Charbonneau \& MacGregor 2001]{cm01} Charbonneau, P., \&
  MacGregor, K. B. 2001, \apj, 559, 1094
  
\bibitem[Cohen, Cassinelli, \& MacFarlane 1997a]{ccm97} Cohen, D. H.,
  Cassinelli, J. P., \& MacFarlane, J. J. 1997, \apj, 487, 867
  
\bibitem[Cohen, Cassinelli, \& Waldron 1997b]{ccw97} Cohen, D. H.,
  Cassinelli, J. P., \& Waldron, W. L. 1997, \apj, 488, 397
  
\bibitem[Donati \etal\ 2002]{d02} Donati, J.-F., Babel, J., Harries,
  T. J., Howarth, I. D., Petit, P., \& Semel, M. 2002, \mnras, 333, 55
  
\bibitem[Donati \etal\ 2001]{d01} Donati, J.-F., Wade, G. A., Babel,
  J., Henrichs, H. F., de Jong, J. A., \& Harries, T. J. 2001, \mnras,
  326, 1265
  
\bibitem[Feldmeier et al. 1997]{f97} Feldmeier, A., Kudritzki, R.-P.,
  Palsa, R., Pauldrach, A. W. A., \& Puls, J. 1997, \aap, 320, 899
  
\bibitem[Feldmeier, Puls, \& Pauldrach 1997]{fpp97} Feldmeier, A.,
  Puls, J., \& Pauldrach, A. W. A. 1997, \aap, 322, 878
  
\bibitem[Gabriel \& Jordan 1969]{gj69} Gabriel, A. H., \& Jordan, C.
  1969, \mnras, 145, 241
  
\bibitem[Harnden \etal\ 1979]{h79} Harnden, F. R., Jr. \etal\ 1979,
  \apj, 234, L51
  
\bibitem[Henrichs \etal\ 2000]{henrichs00} Henrichs, H. F. \etal\ 
  2000, {\it Magnetic Fields of Chemically Peculiar and Related
    Stars}, eds. Yu.V. Glagolevskij, I.I. Romanyuk, 57
  
\bibitem[Howk \etal\ 2000]{h00} Howk, J. C., Cassinelli, J. P.,
  Bjorkman, J. E., \& Lamers, H. J. G. L. M. 2000, \apj, 534, 348
  
\bibitem[Kahn et al. 2001]{k01} Kahn, S. M., Leutenegger, M. A.,
  Cotam, J., Rauw, G., Vreux, J.-M., den Boggende, A. J. F., Mewe, R.,
  \& G\"udel, M.  2001, \aap, 365, L312

\bibitem[Kilian 1992]{k92} Kilian, J. 1992, \aap, 262, 171

\bibitem[Kilian 1994]{k94} Kilian, J. 1994, \aap, 282, 867
  
\bibitem[Lamers \& Rogerson 1978]{lr78} Lamers, H. J. G. L. M., \&
  Rogerson, J. B., Jr. 1978, \aap, 66, 417
  
\bibitem[Linsky \& Gagne 2001]{lg01} Linsky, J. L., \& Gagn\'{e}, M.
  2001, BAAS, 198, 4405
  
\bibitem[Lucy \& Solomon 1970]{ls70} Lucy, L. B., \& Solomon, P. M.
  1970, \apj, 159, 879
  
\bibitem[MacFarlane, Cohen, \& Wang 1994]{mcw94} MacFarlane, J. J.,
  Cohen, D. H., \& Wang, P. 1994, \apj, 437, 351

\bibitem[MacFarlane \etal\ 1993]{m93} MacFarlane, J. J., Wang, P.,
  Bailey, J. E., Melhorn, T. A., Dukart, R. J., \& Mancini, R. 1993,
  \pre, 47, 2748
  
\bibitem[MacGregor \& Cassinelli 2002]{mc-2} MacGregor, K. B., \&
  Cassinelli, J. P. 2002, \apj, in press
  
\bibitem[Mauche, Liedahl, \& Fournier 2001]{mlf01} Mauche, C. W.,
  Liedahl, D. A., \& Fournier, K. B. 2001, \apj, 560, 992
  
\bibitem[Mewe, Kaastra, \& Liedahl 1995]{mkl95} Mewe, R., Kaastra, J.
  S., \& Liedahl, D. A. 1995, Legacy, 6, 16

\bibitem[Miller \etal\ 2002]{m02} Miller, N. A., Cassinelli, J. P.,
  Waldron, W. L., MacFarlane, J. J., \& Cohen, D. H. 2002, \apj, 577, 951
  
\bibitem[Montmerle \etal\ 2002]{mont02} Montmerle, T. Grosso, N.,
  Feigelson, E. D., \& Townsley, L. 2002, in New Visions of the X-ray
  Universe in the {\it XMM-Newton} and {\it Chandra} Era, eds. F.
  Jansen (ESA SP-488; Noordwijk: ESA)

\bibitem[Oskinova \etal\ 2001]{o01} Oskinova, L. M., Ignace, R.,
  Brown, J. C., \& Cassinelli, J. P. 2001, \aap, 373, 1009
  
\bibitem[Owocki, Castor, \& Rybicki 1988]{ocr88} Owocki, S. P.,
  Castor, J. I., \& Rybicki, G. B.  1988, \apj, 335, 914
  
\bibitem[Perryman \etal\ 1997]{p97} Perryman, M. A. C., \etal\ 1997,
  \aap, 323, L49
  
\bibitem[Pradhan \& Shull 1981]{ps81} Pradhan, A. K., \& Shull, J. M.
  1981, \apj, 249, 821
  
\bibitem[Schulz et al.  2001]{s01} Schulz, N. S., Canizares, C. R.,
  Huenemoerder, D., \& Lee, J. C. 2001, \apjl, 549, 441.
  
\bibitem[Smith \etal\ 2001]{aped} Smith, R. K., Brickhouse, N. S.,
  Liedahl, D. A., \& Raymond, J. C. 2001, \apj, 556, L91
  
\bibitem[Smith \& Karp 1979]{sk79} Smith, M. A., \& Karp, A. H. 1979,
  \apj, 230, 156
  
\bibitem[ud-Doula \& Owocki 2002]{uo02} ud-Doula, A., \& Owocki, S. P.
  2002, \apj, 576, 413
  
\bibitem[Walborn, Parker, Nichols 1995]{wpn} Walborn, N. R., Parker,
  J. W., \& Nichols, J. S. 1995, VizieR On-Line Data Catalog, 3188

\bibitem[Waldron \& Cassinelli 2001]{wc01} Waldron, W. L., \&
  Cassinelli, J. P. 2001, \apjl, 548, L45
  
\bibitem[Wolf, Edwards, Preston 1982]{wep82} Wolff, S. C., Edwards,
  S., \& Preston, G. W. 1982, \apj, 252, 322
  
\bibitem[Zaal \etal\ 1999]{z99} Zaal, P. A., de Koter, A., Waters, L.
  B. F. M., Marlborough, J. M., Geballe, T. R., Oliveira, J. M., \&
  Foing, B. H. 1999, \aap, 349, 573

\end{thebibliography}
\end{document}